\documentclass[preprint,prb,aps,footinbib,amsmath,amssymb,superscriptaddress,showkeys,showpacs]{revtex4-1}
\usepackage{graphicx}
\usepackage{braket}
\usepackage{dcolumn}
\usepackage{color}
\usepackage{bm}
\usepackage{natbib,hyperref}
\usepackage{amsmath,amssymb,amsfonts,bbold}
\usepackage[normalem]{ulem}
\usepackage{svg}
\usepackage{listings}
\usepackage[caption=false]{subfig}
\usepackage[usestackEOL]{stackengine}
\usepackage{soul}
\usepackage{gensymb}
\usepackage{mathrsfs}
\usepackage[utf8]{inputenc}
\UseRawInputEncoding 

\makeatletter

\def\revtex@split@authornotes#1{%
  \begingroup
    \def\@separator{;}%
    \protected@edef\x{%
      \endgroup
      \noexpand\revtex@split@authornotes@expanded{#1}%
    }%
  \x
}

\def\revtex@split@authornotes@expanded#1{%
  \@if@empty{#1}{}{%
    \revtex@split@authornotes@i#1;\@nil
  }%
}

\def\revtex@split@authornotes@i#1;#2\@nil{%
  \frontmatter@footnote{#1}%
  \def\revtex@remainingnotes{#2}%
  \@ifx{\revtex@remainingnotes\@empty}{}{%
    \expandafter\revtex@split@authornotes@i#2\@nil
  }%
}

\def\doauthor#1#2#3{%
  \ignorespaces#1\unskip\@listcomma
  \begingroup
    #3%
  \@if@empty{#2}
    {\endgroup{}{}}
    {\endgroup{\comma@space}{}\revtex@split@authornotes{#2}}%
  \space \@listand
}

\makeatother

\newlength{\subcolumnwidth}

\newcommand{\nextsubcolumn}[1][]{%
  \cr\noalign{\hfill}
  \if\relax\detokenize{#1}\relax\else\hsize=#1\setlength{\subcolumnwidth}{\hsize}\fi
}

\newcommand{\ourtitle}{All-electrical Coherent Control of a Single Rare-earth Spin Qubit}

\begin{document}

\title{\ourtitle}

\author{Yaowu Liu}
\thanks{These authors contributed equally to this work.}
\email{liu.yaowu@qns.science}
\affiliation{Center for Quantum Nanoscience (QNS), Institute for Basic Science (IBS), Seoul 03760, Republic of Korea}
\affiliation{Ewha Womans University, Seoul 03760, Republic of Korea}

\author{Dasom Choi}
\thanks{These authors contributed equally to this work.}
\affiliation{Center for Quantum Nanoscience (QNS), Institute for Basic Science (IBS), Seoul 03760, Republic of Korea}
\affiliation{Department of Physics, Ewha Womans University, Seoul 03760, Republic of Korea}

\author{Stefano Reale}
\thanks{These authors contributed equally to this work.}
\affiliation{Center for Quantum Nanoscience (QNS), Institute for Basic Science (IBS), Seoul 03760, Republic of Korea}
\affiliation{Ewha Womans University, Seoul 03760, Republic of Korea}

\author{Jeongmin Oh}
\affiliation{Center for Quantum Nanoscience (QNS), Institute for Basic Science (IBS), Seoul 03760, Republic of Korea}
\affiliation{Department of Physics, Ewha Womans University, Seoul 03760, Republic of Korea}

\author{Seorhin Choi}
\affiliation{Center for Quantum Nanoscience (QNS), Institute for Basic Science (IBS), Seoul 03760, Republic of Korea}
\affiliation{Department of Physics, Ewha Womans University, Seoul 03760, Republic of Korea}

\author{Lei Fang}
\affiliation{Center for Quantum Nanoscience (QNS), Institute for Basic Science (IBS), Seoul 03760, Republic of Korea}
\affiliation{Ewha Womans University, Seoul 03760, Republic of Korea}

\author{We-hyo Soe}
\affiliation{Center for Quantum Nanoscience (QNS), Institute for Basic Science (IBS), Seoul 03760, Republic of Korea}
\affiliation{Ewha Womans University, Seoul 03760, Republic of Korea}

\author{Arzhang Ardavan}
\affiliation{Department of Physics, University of Oxford, The Clarendon Laboratory, Oxford OX1 3PU, U.K}

\author{Andreas J. Heinrich}
\affiliation{Center for Quantum Nanoscience (QNS), Institute for Basic Science (IBS), Seoul 03760, Republic of Korea}
\affiliation{Department of Physics, Ewha Womans University, Seoul 03760, Republic of Korea}

\author{Soo-hyon Phark}
\email{phark@qns.science}
\affiliation{Center for Quantum Nanoscience (QNS), Institute for Basic Science (IBS), Seoul 03760, Republic of Korea}
\affiliation{Ewha Womans University, Seoul 03760, Republic of Korea}

\author{Fabio Donati}
\email{donati.fabio@qns.science}
\affiliation{Center for Quantum Nanoscience (QNS), Institute for Basic Science (IBS), Seoul 03760, Republic of Korea}
\affiliation{Department of Physics, Ewha Womans University, Seoul 03760, Republic of Korea}

\begin{abstract}
\noindent
Electrical control of single spin qubits is a major frontier for nanoscale, high-speed, and scalable quantum devices.
Yet, extending it to highly shielded rare-earth 4$f$ electrons remains an experimental challenge across solid-state platforms.
Here we demonstrate all-electrical coherent control of a single Er electron spin, which is exchange-coupled to a nearby Ti atom. 
Scanning tunneling microscopy-based electron spin resonance with three-dimensional magnetic-field control enables comprehensive mapping of the resonance and Rabi frequencies, revealing pronounced anisotropies in both the Er $\mathfrak{g}$-tensor and the Er-Ti exchange interaction.
The electrical modulation of the anisotropic Er-Ti coupling results in an efficient drive of the Er spin, allowing us to achieve near-gigahertz Rabi frequencies -- a ten-fold improvement over the present record for rare-earth spin qubits. 
By establishing anisotropic exchange as a general resource for electrically accessing shielded rare-earth spins, our results open a new route to ultrafast and local control of rare-earth spins in solid-state quantum devices.
\end{abstract}


\maketitle

\section{Introduction}
\noindent
Rare-earth 4$f$-electron spins provide a versatile platform for quantum technologies, including spin qubits~\cite{bertaina2007rare,zhong2015optically,ranvcic2018coherence,ortu2018simultaneous,le2021twenty,hiraishi2025long}, optical quantum memories~\cite{zhong2015optically,hedges2010efficient,ma2021one,ranvcic2018coherence}, and single-photon sources~\cite{de2008solid,dibos2018atomic}.
Among them, erbium (Er) stands out for its optically addressable ground state, making it highly promising as a spin qubit for spin-photon interfaces at telecom wavelengths~\cite{zhong2015optically,ranvcic2018coherence,ortu2018simultaneous,le2021twenty,hiraishi2025long,benelli2002magnetism,reale2024electrically,gupta2023robust,ourari2023indistinguishable,sell2008field}. 
While single-ion optical control of rare-earth qubits has seen remarkable recent progress~\cite{siyushev2014coherent,zhong2018optically,kindem2020control,raha2020optical,chen2020parallel,uysal2025spin,gritsch2025optical}, electrical control offers a scalable route toward dense integration and ultrafast gate operations~\cite{nowack2007coherent,froning2021ultrafast,wang2022ultrafast}.

Despite demonstrations in bulk crystals and molecular systems~\cite{raha2020optical,bertaina2007rare,thiele2014electrically,vincent2012electronic}, achieving electrical coherent control of an individual rare-earth spin remains a fundamental challenge, arising from the strong shielding of the 4$f$ orbitals by outer shells, which severely limits the transduction of external electric fields into effective driving fields of 4$f$ spins. A mediator spin that is both electrically addressable and precisely coupled to the rare-earth atom is therefore required~\cite{reina2025efficient,phark2023electric}, but engineering such a mediator with the required precision is difficult in conventional solid-state and molecular platforms.

Atomic spins on surfaces provide a powerful platform for precisely engineering such mediator interactions~\cite{wolf2024surface,phark2025roadmap,choi2025electron}, offering identical qubits with atom-by-atom addressability~\cite{wang2023atomic} and bottom-up designability via scanning-probe control~\cite{wang2024construction,yang2021probing}, as well as tunability in charge and spin states~\cite{phark2026spin,czap2025direct}. 
When combining scanning tunneling microscopy with electron spin resonance (ESR-STM), this platform further enables coherent manipulation and readout of a single spin with atomic precision~\cite{yang2019coherent,wang2023atomic,phark2023electric,reina2025efficient}.
Despite recent advances in the readout of single rare-earth atoms, for example, by exploiting partially filled 6$s$ shells~\cite{czap2025direct} or coupling to a nearby sensor spin~\cite{reale2024electrically}, coherent electrical manipulation of an individual rare-earth spin still remains elusive.

Here we exploit a tailored Er-Ti spin pair on an MgO/Ag(100) surface, where the contrasting $\mathfrak{g}$-tensor anisotropies of Er and Ti cause the two coupled spins to respond differently to the oscillating electric field, generating an effective transverse driving field on the Er spin tunable by the magnetic field direction. We achieve a near-gigahertz Rabi frequency, a record for rare-earth spin qubits, and identify a general mechanism for electrical coherent control based on the effective wobbling of coupled magnetic moments.

\section{Remote drive of a single \NoCaseChange{Er} spin}
\begin{figure}
    \centering
    \includegraphics[width = 1
\textwidth]{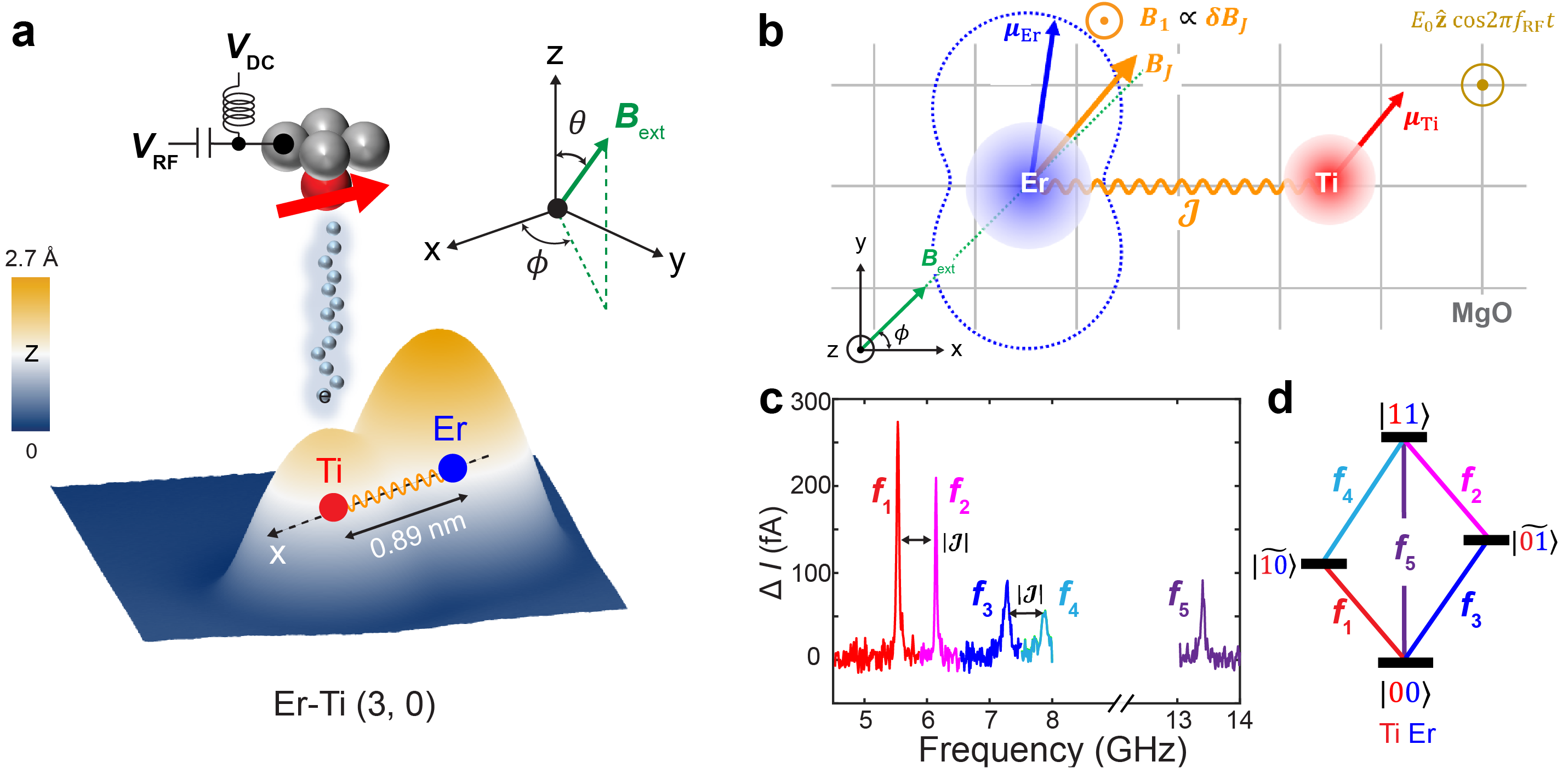}
    \caption{\textbf{Schematics and ESR spectrum of an Er-Ti pair.}
    \textbf{a,} Schematic and three-dimensional STM topograph of an Er-Ti pair oriented along the O-O direction on MgO, $3.0~{\rm nm}\times 3.0~{\rm nm}$. The gray cylinder indicates the tunneling current, with the small gray balls representing the tunneling electrons.
    \textbf{b,}  Schematic of an Er-Ti spin pair along the O-O direction on MgO, coupled by the spin-spin interaction $\mathcal{J}$. Owing to the anisotropic $\mathfrak{g}$-tensors, the magnetic moments of Ti and Er, $\boldsymbol{\mu}_{\rm Ti}$ and $\boldsymbol{\mu}_{\rm Er}$, are generally noncollinear with $\boldsymbol{B}_{\rm ext}$. The peanut-shaped blue dashed line shows a representative anisotropic magnetic moment of Er for the situation of $\mathfrak{g}^{\rm Er}_{yy}\sim 2\mathfrak{g}^{\rm Er}_{xx}$. The nearby Ti spin generates an effective exchange field ($\boldsymbol{B}_{\mathcal{J}}$) acting on Er; its RF-induced modulation provides a transverse driving field, $\boldsymbol{B}_{1}\propto\delta\boldsymbol{B}_{\mathcal{J}}$, for Er spin transitions. The RF electric field is applied along the $z$-axis. The gray lines indicate the MgO lattice, with intersections marking oxygen atomic sites. The surface lattice constant is defined by the nearest-neighbor O-O distance.
   \textbf{c,} Representative ESR spectrum of the Er-Ti pair measured at $\boldsymbol{B}_{\rm ext}=0.25~{\rm T}$, $\theta=83\degree$, and $\phi=52\degree$. 
\textbf{d,} Energy-level diagram of the Er-Ti spin pair. Unless otherwise noted, all data were taken at $V_{\rm DC} = 100~{\rm mV}$, $I_{\rm DC} = 20~{\rm pA}$, and $T = 1.0~{\rm K}$. The data in (\textbf{c}) were acquired with $V_{\rm RF}=20~{\rm mV}$.}
\label{Fig1}
\end{figure}

\noindent
Here we implement interaction-mediated driving of a single rare-earth spin, in which electrical excitation of a neighboring spin is transduced into an effective transverse magnetic field acting on the target spin. To realize this concept, we engineer a tailored Er-Ti spin pair on an MgO/Ag(100) surface (Methods), as shown in the scanning tunneling microscopy (STM) topography image in Fig.~\ref{Fig1}\textbf{a} and schematic in Fig.~\ref{Fig1}\textbf{b}. 
The design combines an Er atom occupying a low-symmetry binding site on the MgO surface, namely an oxygen-oxygen bridge site [Er(B)], with exchange coupling $\mathcal{J}$ to a nearby Ti atom.
This low-symmetry binding site gives Er a strongly anisotropic magnetic moment ($\boldsymbol{\mu}_{\rm Er}$, Fig.~\ref{Fig1}\textbf{b}). 
As a result, the Ti spin generates an effective exchange field on Er ($\boldsymbol{B}_{\mathcal{J}}\propto \boldsymbol{\mu}_{\rm Ti}\mathcal{J}$). Under RF electric excitation, modulation of this exchange field produces a transverse driving field ($\boldsymbol{B}_1\propto \delta\boldsymbol{B}_{\mathcal{J}}$), enabling coherent control of the Er spin (Fig.~\ref{Fig1}\textbf{b}).
The pair is assembled by atom manipulation~\cite{eigler1990positioning} (Fig.~\ref{Fig1}\textbf{a}), with Er and Ti separated by three surface lattice sites along the oxygen-to-oxygen (O-O) direction. Taking advantage of this geometry, we define the $x$-axis along the O-O direction, the $y$-axis perpendicular to it, and the $z$-axis along the surface normal, corresponding to the pair geometry $(3,0)$, as shown in Fig.~\ref{Fig1}\textbf{a}.

Erbium adsorbed at a bridge site of the MgO surface, denoted hereafter as Er unless otherwise specified, shows neither spin excitations in tunneling spectroscopy nor ESR signals when addressed directly (Supplementary Section I). This behavior is consistent with the absence of unpaired 5$d$ and 6$s$ electrons, similar to Er atoms adsorbed at oxygen sites [Er(O)]~\cite{reale2024electrically}.
In contrast, when the STM tip is positioned over the Ti atom of the Er-Ti pair and an RF electric field is applied, five ESR transitions are observed, as shown in Fig.~\ref{Fig1}\textbf{c}. 
The four lower frequency peaks can be divided into two groups ($f_1$, $f_2$ and $f_3$, $f_4$), showing the same splitting ($f_2-f_1$ and $f_4-f_3$) of a few hundred MHz, which reflects the Er-Ti spin-spin interaction strength. 
The relative amplitudes of the two transitions in each group reveal an antiferromagnetic spin-spin coupling~\cite{yang2017engineering}. 
Similar to the Er(O)-Ti case, the eigenstates of the Er(B)-Ti spin pair form a four-level manifold: $\ket{00}$, $\ket{11}$, $\ket{+}=\cos(\eta/2)\ket{01}+\sin(\eta/2)\ket{10}$, and $\ket{-}=\sin(\eta/2)\ket{01}-\cos(\eta/2)\ket{10}$, where $\tan(\eta)=|\mathcal{J}|/\delta$ and $\delta$ is the Zeeman-energy detuning between the two spins.
We ascribe the above four ESR peaks to the four single quantum transitions (SQT, $f_1$ to $f_4$, Fig.~\ref{Fig1}\textbf{d}). The fifth transition ($f_5$) appears at twice the mean frequency of the four single quantum transitions (Fig.~\ref{Fig1}\textbf{d} and Supplementary Section II), and it is related to a double quantum transition (DQT) in which both Er and Ti spins are rotated simultaneously~\cite{reale2024electrically}.

\section{Fast coherent control of $4f$ electrons}
\noindent
First, we applied a nearly in-plane magnetic field ($\textbf{\textit{B}}_{\rm ext}$ = 0.25 T, $\theta$ = 83$\degree$) at an  azimuthal angle of $\phi$ = 52$\degree$ that maximizes the intensity of the single-quantum transitions (Fig.~\ref{Fig1}\textbf{c}). Under this field orientation, the system eigenstates are tuned to be approximately product states in the Zeeman basis (Supplementary Section IV): $\ket{+}\approx {\ket{\widetilde{01}}}$ and $\ket{-}\approx {\ket{\widetilde{10}}}$, as shown in Fig.~\ref{Fig1}\textbf{d}.
The response of each peak to the magnetic field from the spin-polarized tip reveals that the transitions $f_3$ and $f_1$ are dominated by the Er and Ti spin resonances, respectively (Supplementary Section V).

\begin{figure}
    \centering
    \includegraphics[width = 1\textwidth]{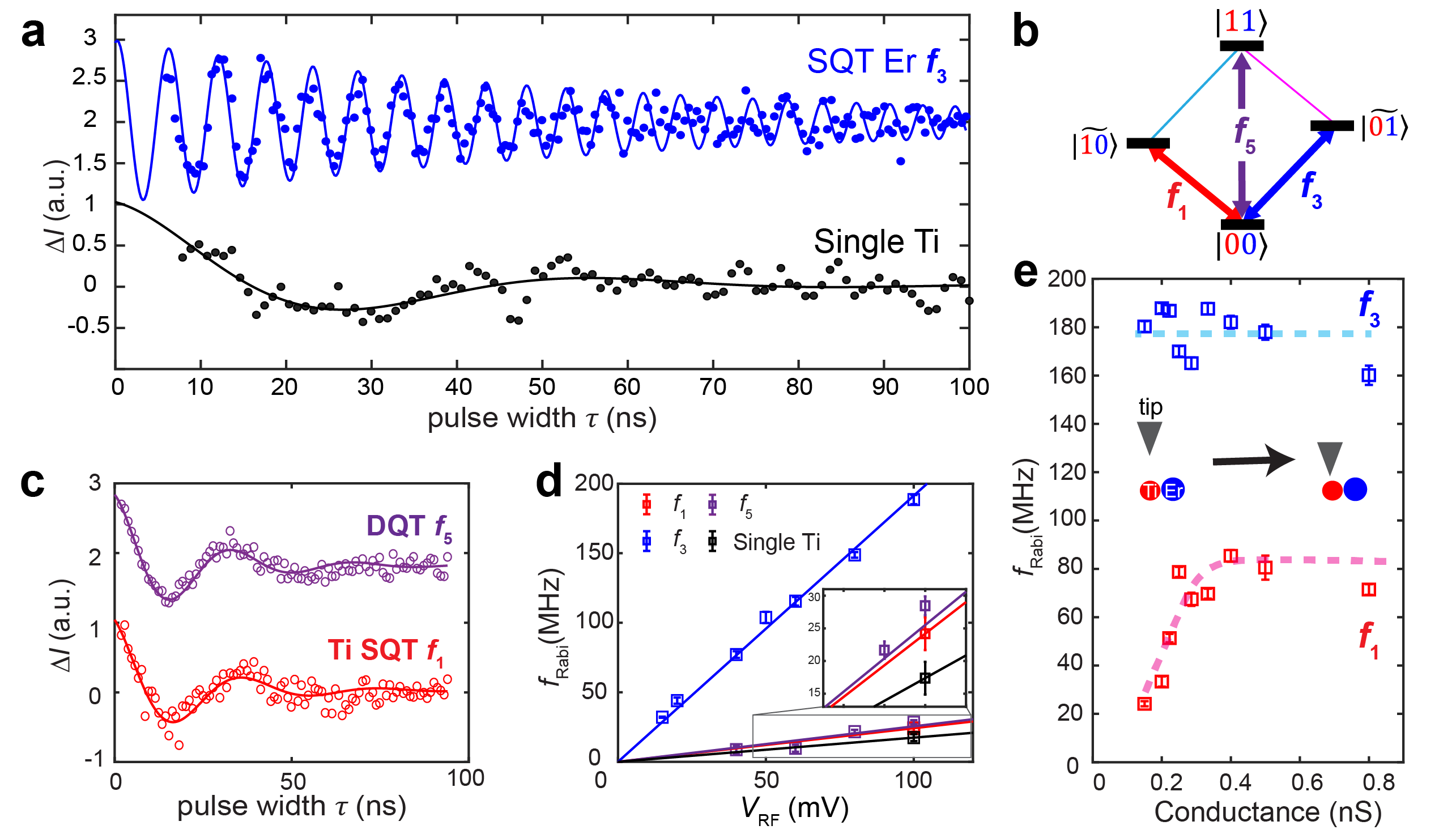}
    \caption{\textbf{Coherent control of the Er spin.} \textbf{a,} Rabi oscillations of the $f_3$ transition ($f_{\rm res}$ = 7.28 GHz) in the pair and in a single Ti spin on the same surface. Both datasets were taken using the same tip with $V_{\rm DC}$ = 100 mV, $I_{\rm DC}$ = 20 pA. The solid lines are fits to exponentially decaying sinusoids. \textbf{b,}  Energy diagram under this field condition where $\ket{\pm}$ are approximately Zeeman product states: $\ket{\widetilde{01}}$ and $\ket{\widetilde{10}}$. \textbf{c,} Rabi oscillations of $f_1$ and $f_5$ with the solid lines showing fits to exponentially decaying sinusoids. \textbf{d,} Rabi frequency as a function of $V_{\rm RF}$. Inset shows a magnified view of Rabi frequency vs $V_{\rm RF}$ specifically for $f_1$ ($f_{\rm res}$ = 5.55 GHz) and $f_5$ ($f_{\rm res}$ = 13.44 GHz). \textbf{e,}  Rabi frequency of $f_3$ (blue squares) and $f_1$ (red squares) as a function of the conductance. The dashed cyan lines are guides to the eye. All data except in (\textbf{e}) were taken with a junction condition of $V_{\rm DC}$ = 100 mV, $I_{\rm DC}$ = 20 pA, $\boldsymbol{B}_{\rm ext}$ = 0.25 T, $\theta$ = 83$\degree$, $\phi$ = 52$\degree$. All data were acquired at 1.0 K.}
    \label{Fig2}
\end{figure}

We performed Rabi measurements on the transition $f_3$,
yielding coherent oscillations (Fig.~\ref{Fig2}\textbf{a}) with a Rabi frequency of 190 MHz, which is one order of magnitude higher than the previously reported record for Er-based qubits in solid-state platforms~\cite{weiss2025high}.
In stark contrast, the Rabi sequence measured on an isolated Ti atom shows only weak oscillations (Fig.~\ref{Fig2}\textbf{a}) with a Rabi frequency of $19~\mathrm{MHz}$, an order of magnitude lower than that of the Er transition.
Such a high Rabi frequency is observed exclusively from the Er spin transition ($f_3$) and not the case for the Ti transition ($f_{1}$) or the DQT ($f_5$) in the pair (Figs.~\ref{Fig2}\textbf{c} and \textbf{d}, red and purple squares), which remain similar to that of the isolated Ti atom under the same conditions. 
The Rabi frequencies for all transitions scale linearly with the applied RF voltage $V_{\rm RF}$ (Fig.~\ref{Fig2}\textbf{d}), confirming that the oscillations arise from coherent RF-driven spin rotations. Notably, the slope for the Er transition $f_3$ is about an order of magnitude larger than those of the Ti SQT and DQT, highlighting the strongly enhanced driving efficiency of the Er $4f$ spin.
This enhanced driving efficiency identifies the Er 4$f$ spin as a compelling platform for electrically controlled spin qubits on surfaces.

A further interesting feature is the independence of the Er $f_3$ Rabi frequencies on the tunneling conductance (Fig.~\ref{Fig2}\textbf{e}), i.e., the tip-Ti distance, excluding the role of $\textbf{\textit{B}}_{\rm tip}$ in the coherent driving of the Er spin.
In contrast, the Rabi frequency of the Ti transition $f_1$ in the pair initially increases with decreasing tip-Ti distance, consistent with an enhanced tip-induced driving field~\cite{phark2023electric}, and then saturates (Fig.~\ref{Fig2}\textbf{e}, red squares; see also Supplementary Section VI).

We note that coherent control of the DQT is of particular interest for applications in quantum entanglement~\cite{dolde2013room,morillas2025entanglement} and quantum sensing~\cite{barry2020sensitivity}, but has not been demonstrated previously for surface atoms. 
The DQT requires the coherent rotation of both Er and Ti spins, which suggests the coherent control mechanism is related to the spin-spin interaction, as further discussed in the following section. 

\begin{figure*}
    \centering
    \includegraphics[width = 0.6\textwidth]{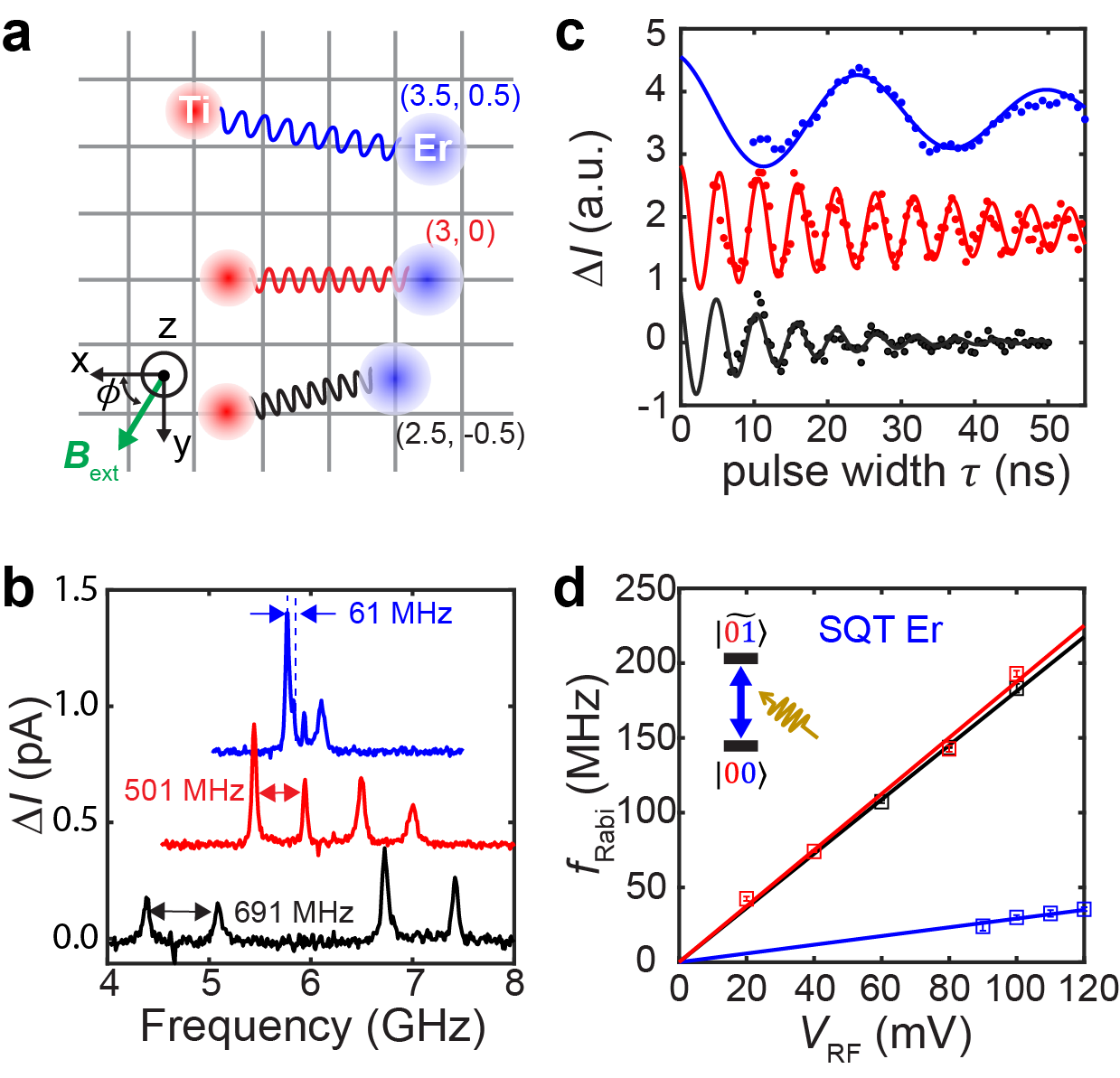}
    \caption{\textbf{Pair engineering.} \textbf{a,} Schematics for three different pairs. The gray grid represents the MgO lattice with oxygen sites at the intersections. \textbf{b,} ESR spectra of the three pairs taken at $\textbf{\textit{B}}_{\rm ext}$ = 0.25 T, $\phi$ = 52$\degree$ and  $\theta$ = 90$\degree$. The interaction increases from 61 MHz to 691 MHz with decreasing separation between two spins. \textbf{c,} Rabi oscillations of $f_3$ (Er transition) in each pair  ($V_{\rm RF}$ = 100 mV, $f_{\rm res}$ = 6.10  GHz for (3.5, 0.5) pair, $f_{\rm res}$ = 6.40 GHz for (3, 0) pair, and $f_{\rm res}$ = 4.38 GHz for (2.5, $-$0.5) pair). 
    \textbf{d,} Rabi frequency of $f_3$ in each pair as a function of $V_{\rm RF}$. All data were taken with $V_{\rm DC}$ = 100 mV and $I_{\rm DC}$ = 20 pA at 1.0 K.}
    \label{Fig3}
\end{figure*}

To survey the influence of the spin pair geometry, i.e., atomic separation and pair orientation, on the Rabi frequency, we prepared two more distinct Er-Ti pairs of surface lattice coordinate configurations, (3.5, 0.5) and (2.5, $-$0.5), as depicted in Fig.~\ref{Fig3}\textbf{a}, and performed ESR measurements for a comparative study with the (3, 0) pair.
ESR spectra taken at the same magnetic field reveal a clear monotonic increase in the splitting between the transitions $f_1$ and $f_2$ with decreasing separation (Fig.~\ref{Fig3}\textbf{b}, colored arrows), consistent with the expected distance dependence of the spin-spin interaction.
By contrast, the Rabi frequencies of the Er transitions show an interesting behavior (Figs.~\ref{Fig3}\textbf{c} and \textbf{d}), markedly increasing from (3.5, 0.5) to (3, 0), however does not show a particular change by a further decrease of separation. 
This suggests a nontrivial correlation between spin-spin interaction and driving strength, which will be explained in the following section.
The coherence times of the Er transitions in these pairs are on the order of $100~\mathrm{ns}$, possibly limited by the lifetime of the Ti spin~\cite{wang2023atomic} (Supplementary Section VII).

\section{Magnetic anisotropy of \NoCaseChange{Er-Ti} spin pair}
\begin{figure}
    \centering
    \includegraphics[width = 0.450\textwidth]{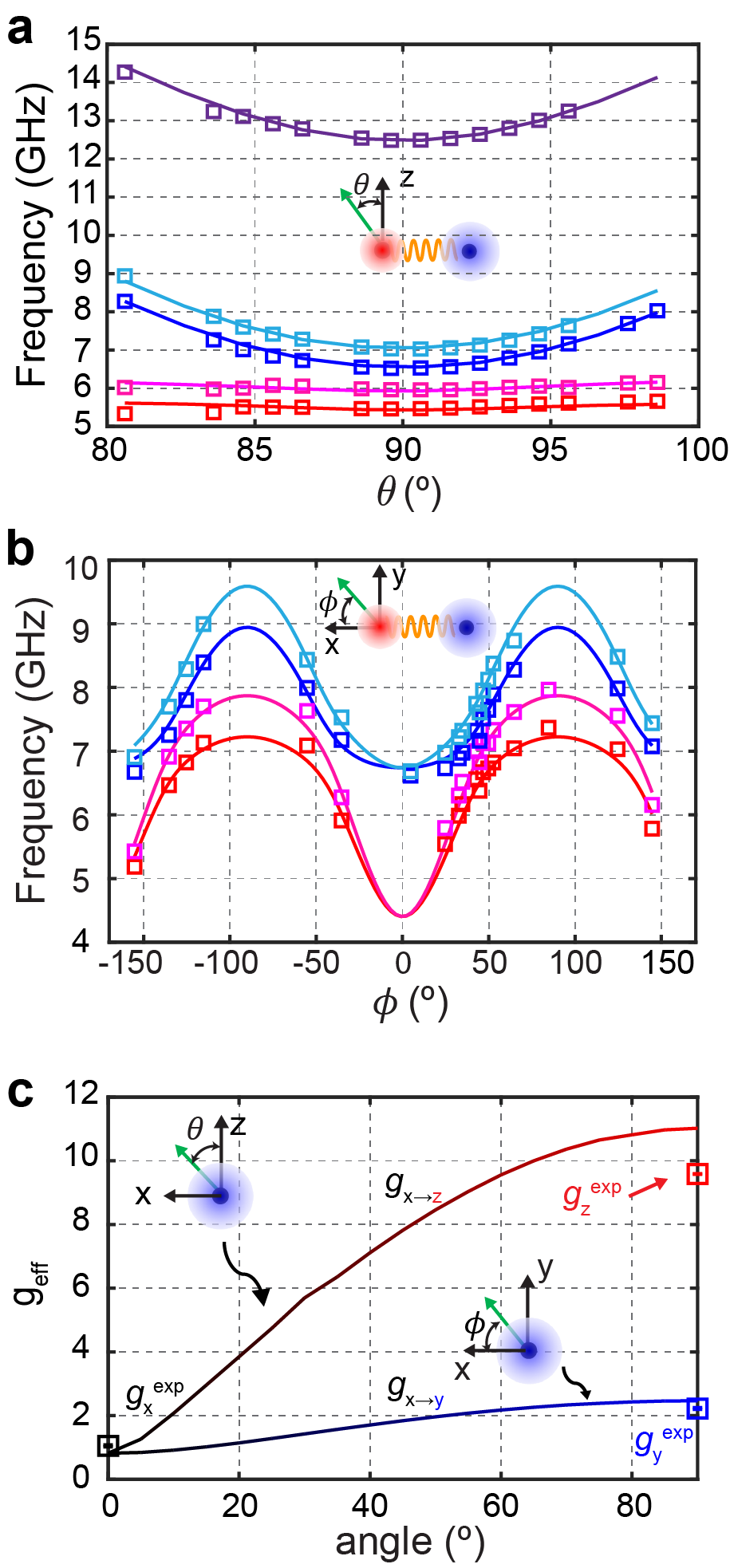}
    \caption{\textbf{Magnetic anisotropy of the Er-Ti (3, 0) pair.} \textbf{a,} ESR transition energy as a function of the polar angle $\theta$ of $\boldsymbol{B}_{\rm ext}$ with $\phi$ fixed at $52\degree$. \textbf{b,} ESR transition energy as a function of the azimuthal angle $\phi$ of $\textbf{\textit{B}}_{\rm ext}$ ($\theta = 90 \degree$). Data in (\textbf{a}) were taken with $V_{\rm RF}$ = 20 mV, $V_{\rm DC}$ = 100 mV, $I_{\rm DC}$ = 20 pA and $\boldsymbol{B}_{\rm ext} = 0.25~{\rm T}$. Data in (\textbf{b}) were taken with $V_{\rm RF}$ = 20 mV, $V_{\rm DC}$ = 50 mV, $I_{\rm DC}$ = 20 pA and $\boldsymbol{B}_{\rm ext} = 0.3~{\rm T}$. \textbf{c,} The principal values of $\mathfrak{g}$-tensor from multiplet calculation (solid lines) and fitting of experimental data (colored squares). The solid lines show how the principal values of the $\mathfrak{g}$-tensor change as a function of $\theta$ and $\phi$ (see insets). All data were taken at 1.0 K. }
    \label{Fig4}
\end{figure}

\noindent
From the data presented in Figs.~\ref{Fig2} and \ref{Fig3}, it is clear that the effective transverse driving field $\boldsymbol{B}_1$ acting on the Er spin does not originate from the tip field, and it is not only related to the magnitude of the spin-spin interaction.
To provide a detailed model, we reconstructed the static Hamiltonian for the Er-Ti (3, 0) pair by taking advantage of our three-dimensional control of the magnetic field at a fixed magnitude.
We first measured the ESR transition energies as a function of magnetic-field polar angle $\theta$
while the azimuthal angle $\phi$ was fixed at 52$\degree$ (Fig.~\ref{Fig4}\textbf{a}). 
The first transition group ($f_1$, $f_2$) remains nearly unchanged at around $6~\mathrm{GHz}$, reflecting almost isotropic couplings (i.e. $\mathfrak{g}$-tensor) of the spins to the external field in this field-rotation plane.
The weak anisotropy in these transition energies suggests that these transitions are mainly related to the Ti spin~\cite{yang2019coherent,yang2017engineering,seifert2020longitudinal}, confirming the results deduced from the tip-field dependence. 
By contrast, the second group ($f_3$, $f_4$) shows a substantial increase in the transition energy, indicating a significantly anisotropic $\mathfrak{g}$-tensor of the spin that dominates these transitions. 
Their intensities decrease when the field rotates away from the in-plane direction ($\theta = 90\degree$) and eventually smear out with the field along the surface normal (see Fig.~\ref{Fig5}\textbf{d}), suggesting the driving is significantly influenced by the field direction. 

Similarly, rotating the magnetic field within the surface plane reveals a pronounced angular dependence of the ESR spectra, with all four transitions exhibiting substantial shifts in frequency across the range of 5 - 9 GHz (Fig.~\ref{Fig4}\textbf{b}). This behavior indicates a strong in-plane $\mathfrak{g}$-tensor anisotropy of Er and highlights the tunable eigenstates using the field orientation. Notably, when the field is aligned with the principal crystallographic axes ($\phi = 0\degree$ and $90\degree$), the transitions $f_3$ and $f_4$ become completely unobservable (Fig.~S16), reflecting a pronounced angular selectivity of the allowed ESR transitions.
Of particular interest, the interaction strength between the two spins, extracted from the splitting between $f_1$ and $f_2$ (or $f_3$ and $f_4$, Fig.~\ref{Fig1}\textbf{c}), shows a pronounced change with the angle $\phi$.
This indicates an anisotropic exchange interaction, which, to our knowledge, has not previously been observed in other on-surface spin systems.

To interpret these experimental observations, we map the Er ground-state doublet onto an effective spin of $S=1/2$ ($\boldsymbol{S}_{\rm Er}$) and construct a spin Hamiltonian for the Er-Ti pair:
\begin{equation}
    \label{eq.1}
   \mathcal{H} = 
   -\mu_{\rm B}\boldsymbol{S}_{\rm Ti} \cdot \mathfrak{g}^{\rm Ti}\cdot(\boldsymbol{B}_{\rm ext}+\boldsymbol{B}_{\rm tip}) -\mu_{\rm B}\boldsymbol{S}_{\rm Er} \cdot \mathfrak{g}^{\rm Er}\cdot\boldsymbol{B}_{\rm ext} + \boldsymbol{S}_{\rm Ti} \cdot \mathcal{J}\cdot \boldsymbol{S}_{\rm Er},
\end{equation}
where $\mu_{\rm B}$ is the Bohr magneton, $\boldsymbol{B}_{\rm ext}$ is the external magnetic field, $\boldsymbol{B}_{\rm tip}$ is the magnetic field from the spin-polarized tip, $\mathfrak{g}^{\rm Ti}$ and $\mathfrak{g}^{\rm Er}$ are the $\mathfrak{g}$-tensors of Ti and Er, respectively, and $\mathcal{J}$ is the effective interaction tensor, including both exchange ($\mathcal{J}^{\rm exc}$) and dipolar contributions (Supplementary Section III).
As in previous works, the effective tip field $\boldsymbol{B}_{\rm tip}$ only applies to the spin under the tip, in this case Ti~\cite{yang2017engineering}.

Considering the symmetry of the Er-Ti pair given by the adsorption sites, only diagonal components of $\mathfrak{g}^{\rm Er}$ and $\mathcal{J}$ are non-zero. The experimental data, including transition intensities and Rabi frequencies described later, are well reproduced by taking $\textbf{\textit{B}}_{\rm tip}$ as a free parameter (solid curves in Figs.~\ref{Fig4}\textbf{a} and \textbf{b}; Supplementary  Section III). 
Using the known $\mathfrak{g}^{\rm Ti}$ tensor~\cite{kim2021spin}, the fit yields $\mathfrak{g}^{\rm Er}_{xx}=1.05\pm0.02$, $\mathfrak{g}^{\rm Er}_{yy}=2.22\pm0.02$, and $\mathfrak{g}^{\rm Er}_{zz}=9.59\pm0.04$, revealing a strong anisotropy of the Er(B) magnetic moment with an out-of-plane easy axis (Fig.~\ref{Fig4}\textbf{c}).
The two-fold symmetry of the oxygen-bridge binding site also leads to different $g$ factors along the two in-plane directions, in contrast to the case of Er(O)~\cite{reale2023erbium,reale2024electrically}. 
In addition, the fit yields the exchange components of $\mathcal{J}^{\rm exc}_{xx}$ = 79.6 $\pm$ 5.5 MHz, $\mathcal{J}^{\rm exc}_{yy}$ = 543.7 $\pm$ 6.3 MHz, and $\mathcal{J}^{\rm exc}_{zz}$ = 1272.5 $\pm$ 7.8 MHz.
The pronounced anisotropy in both $\mathcal{J}^{\rm exc}$ and $\mathfrak{g}^{\rm Er}$ originates from the combined effects of strong spin-orbit coupling (SOC) and the crystal-field potential under the reduced symmetry of Er(B)~\cite{vieru2016giant}. These anisotropies are therefore distinct from those of Er(O), owing to the different ligand fields at the two binding sites.
For Er(O), the quantum states are split in an almost uniaxial ligand field provided by the oxygen atom underneath. By contrast, atomic multiplet calculations of Er(B) (Supplementary Section VIII) combined the electrostatic potential of O/Mg atoms together with the occupation of the $5d_{y^2}$ orbital, elongated along the $y$-axis (i.e., perpendicular to O-O direction), to impose an out-of-plane easy axis for Er(B).
The calculated principal values of $\mathfrak{g}$-tensor, as shown in Fig.~\ref{Fig4}\textbf{c} (solid lines) qualitatively agree with the experimental results.

\section{Coherent Control Mechanism}
\noindent
Using the parameters obtained from the fits in Section II.\textbf{B}, we now turn to the mechanism generating the driving field $\boldsymbol{B}_1$ for such large Rabi frequencies shown in Section II.\textbf{A}.
Based on the finding that the $\boldsymbol{B}_1$ acting on the Er spin depends not only on the separation but also on the orientation of an Er-Ti pair (Fig.~\ref{Fig3}), we write the spin-pair interaction $\mathcal{J}(\boldsymbol{q})$ using a generalized vector of three variables $\boldsymbol{q}=(\boldsymbol{r},\boldsymbol{\eta}_{\rm Ti},\boldsymbol{\eta}_{\rm Er})$: $\boldsymbol{r}$, the vector connecting two spins, and $\boldsymbol{\eta}_{\rm Ti(Er)}$, the orientations of Ti (Er) orbitals in the ground state of the pair with respect to the lattice.

\begin{figure}
    \centering
    \includegraphics[width = 1\textwidth]{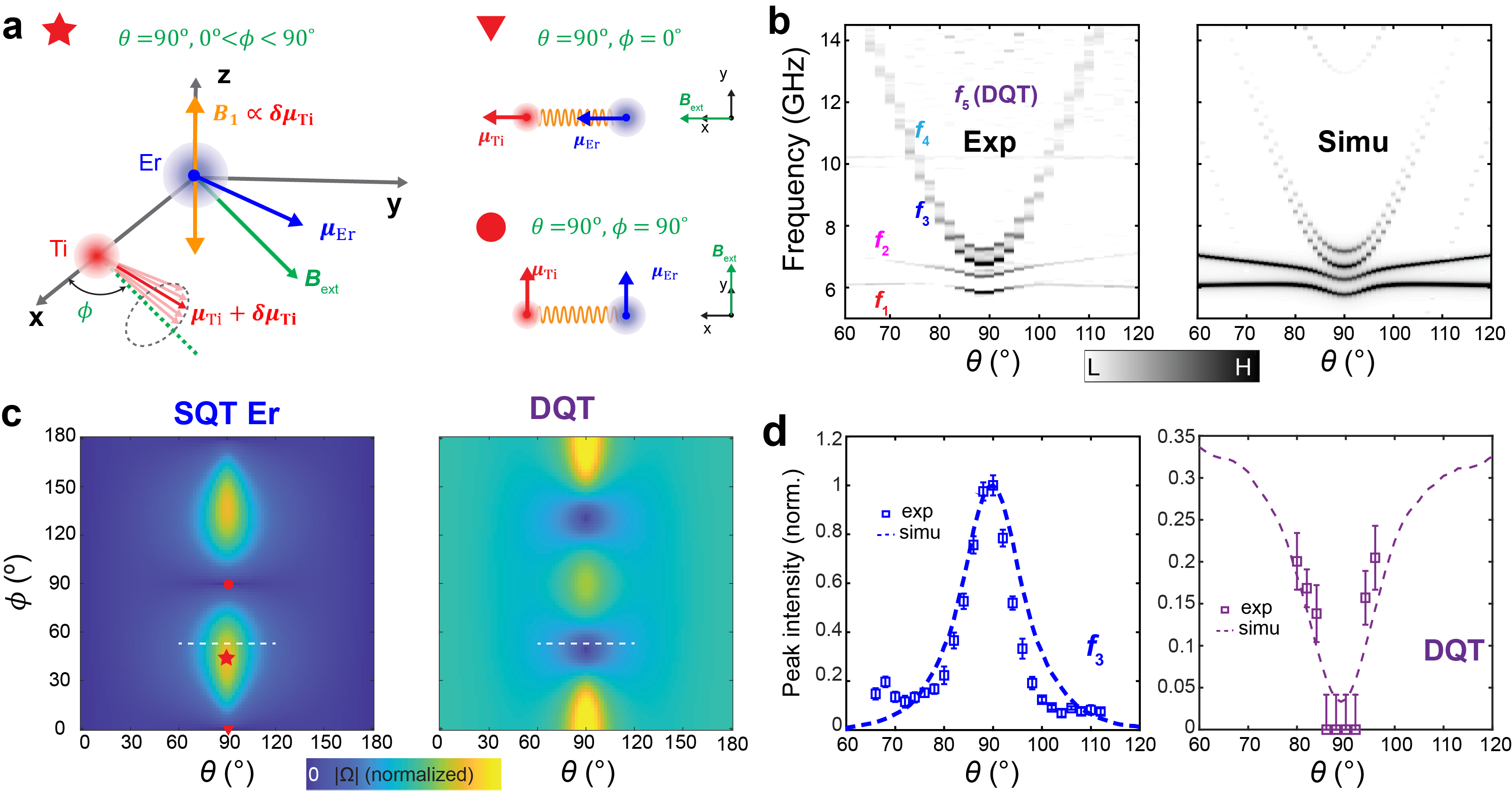}
    \caption{\textbf{Coherent control mechanism.} \textbf{a,} Schematics of coherent control mechanism for special cases of $\theta=90\degree$ and $0\degree<\phi<90\degree$ (left panel), $\theta=90\degree$ and $\phi=0\degree$ (upper-right panel) and $\theta=90\degree$ and $\phi=90\degree$ (lower-right panel), corresponding to the markers in (\textbf{c}). When $\theta=90\degree$ and $0\degree<\phi<90\degree$, the Er and Ti magnetic moments are tilted away from $\boldsymbol{B}_{\rm ext}$ toward the $y$ axis. 
    The wobbling of the Ti magnetic moment around $\boldsymbol{B}_{\rm ext}$ periodically generates an out-of-plane component of the Ti moment, as illustrated in Fig.~\ref{Fig1}\textbf{b}. Through the anisotropic interaction, this produces an out-of-plane effective field, which is strongly amplified by the large $\mathfrak{g}^{\rm Er}_{zz}$ and appears as a transverse driving field for Er spin. 
    When $\theta=90\degree$ and $\phi=0\degree$ or $\phi=90\degree$, the spins are aligned with $\boldsymbol{B}_{\rm ext}$ and therefore the wobbling of magnetic moments does not generate any effective field driving single spin transitions.
    \textbf{b,} Experimental and simulated ESR maps as a function of the polar angle $\theta$ at $\phi=52\degree$ and $\boldsymbol{B}_{\rm ext}=0.25~{\rm T}$. The simulation is based on rotational modulation within a Lindblad formalism (Supplementary Section X). 
    \textbf{c,} Simulated Rabi frequency maps based on rotational modulation for the Er SQT (left) and the DQT (right). \textbf{d,} ESR intensities of $f_3$ and the DQT as a function of the $\theta$. Dashed lines correspond to simulated results in (\textbf{b}) and colored squares denote experimental data. All data were acquired at $T=1.0~{\rm K}$ with $V_{\rm DC}=100~{\rm mV}$, $I_{\rm DC}=20~{\rm pA}$, and $V_{\rm RF}=20~{\rm mV}$. The experimental data in (\textbf{b, d}) correspond to the $\phi$ angle indicated by the dashed white lines in (\textbf{c}).}
    \label{Fig5}
\end{figure}

Our model is as follows: an RF drive results in the time-dependence of $\mathcal{J}$: $\mathcal{J}(\boldsymbol{q}+\delta\boldsymbol{q}(t))$ and generates non-zero off-diagonal components, together with the diagonal components in $\mathcal{J}$ (Supplementary  Section IX).
We first tried to simulate ESR spectroscopy of the spin pair only by modulating the diagonal components, which fails to describe the experimental trends (Figs.~S9-S11).
By contrast, simulations using the off-diagonal terms successfully reproduce all the experimental ESR transitions and the dependence of their intensities on the magnetic field angles (Fig.~\ref{Fig5}\textbf{b}).
In general, $\mathcal{J}(\boldsymbol{q}+\delta\boldsymbol{q}(t))$ depends on the trajectory of $\boldsymbol{q}(t)$ (Supplementary Section IX).
To minimize the energetic cost of the RF excitation, we consider the rotational trajectory that leaves the dominant Zeeman energy unchanged. 
Physically, the RF electric field modulates the orbital structure that defines the principal axes of the $\mathfrak{g}$-tensor and the exchange tensor $\mathcal{J}^{\rm exc}$. 
Through spin-orbit coupling, this modulation can be equivalently described as a wobbling of the magnetic moments around the external field, $\boldsymbol{B}_{\rm ext}$ (Fig.~\ref{Fig5}\textbf{a} and Fig.~S12):
\begin{equation}
    \label{EQ7}
    \mathcal{H}_{1}=\boldsymbol{S}_{\rm Ti}\left[\hat{R}_{\rm B}^T(\delta\alpha)\mathcal{J}\hat{R}_{\rm B}(\delta\alpha)-\mathcal{J}\right]\boldsymbol{S}_{\rm Er}=\delta\alpha\boldsymbol{S}_{\rm Ti}\left[\mathcal{J}, G\right]\boldsymbol{S}_{\rm Er}
\end{equation}
where $\hat{R}_{\rm B}(\delta\alpha)$ is the $SO(3)$ rotational operator around $\boldsymbol{B}_{\rm ext}$; $G$ is the generator of $\hat{R}_{\rm B}(\delta\alpha)$ and $\delta\alpha(t)=\delta\alpha_0\cos{\omega_{\rm RF} t}$ represents the infinitesimal angular modulation of magnetic moments induced by the RF excitation.
Note that this mechanism is only valid for anisotropic $\mathcal{J}$ and a more general formula taking into account the $\mathfrak{g}$-tensor anisotropy is given in Supplementary Section IX and the Methods section.
Using the driving term in Eq.~(\ref{EQ7}), we calculate Rabi frequencies in the spin pair as shown in Fig.~\ref{Fig5}\textbf{c} for the transitions $f_3$ (Er-like transition) and $f_5$ (DQT). Their dependence on the field direction ($\theta$, $\phi$) reveals strong anisotropies. The Rabi frequencies in both transitions are enhanced when an in-plane field is applied, but show an opposite $\phi$-dependence, e.g., maximum for $f_3$ and minimum for $f_5$ at $\phi\sim 50\degree$ (marked by the red star). We further simulate the ESR spectra at $\phi=52\degree$ with a varying $\theta$ for all five transitions and compare with the experimental data (Fig.~\ref{Fig5}\textbf{b}, Supplementary Section X). In Fig.~\ref{Fig5}\textbf{d}, we take a careful look at the $\theta$-dependence of the ESR intensities for $f_3$ and $f_5$, showing excellent agreement between simulations and experiments.

The angular maps in Fig.~\ref{Fig5}\textbf{c} allow us to intuitively grasp the coherent manipulation mechanism in our spin pair: The Rabi frequency vanishes (Fig.~\ref{Fig5}\textbf{c}) along the two in-plane lattice directions ($\theta$ = 90$\degree$; $\phi$ = 0$\degree$ or 90$\degree$), clearly supporting the vanishing ESR intensity in the experiments (Fig.~S16). The magnetic moments of both spins ($\bm{\mu}_{\rm Er}$, $\bm{\mu}_{\rm Ti}$) in this case are fully aligned with the field
(see schemes with triangular and circular markers in Fig.~\ref{Fig5}\textbf{a}), preventing their rotational modulations.
As $\phi$ deviates from 0$\degree$ or 90$\degree$, neither $\bm{\mu}_{\rm Er}$ nor $\bm{\mu}_{\rm Ti}$ is aligned with the external field due to the anisotropies of their $\mathfrak{g}$-tensors. Here we note the much stronger in-plane anisotropy of the Er $\mathfrak{g}$-tensor than that of Ti, leading to deviation of Er quantization axis toward the $y$-direction, which is much more pronounced than that of Ti (Figs.~\ref{Fig1}\textbf{b} and \ref{Fig5}\textbf{a}).
Therefore, the rotational modulation of $\bm{\mu}_{\rm Ti}$ produces an effective transverse field $\boldsymbol{B}_1$ along the out-of-plane direction, which couples to $\mathfrak{g}^{\rm Er}_{\rm zz}$ and generates the Rabi coupling. As we found in Fig.~\ref{Fig4}, the surprisingly large $\mathfrak{g}^{\rm Er}_{\rm zz}$ = 9.59, thus explains the strong Rabi frequency of the Er transition ($f_3$) (also see Fig.~S17). 
However, the mixture of rotation along $x$ and $y$ directions at this angle accidentally depletes the Rabi coupling of the DQT, also in good agreement with the experiments (Fig.~\ref{Fig5}\textbf{c}). 

Our model also reproduces the angular dependence of the SQT, DQT, and the Rabi frequencies, measured from the (2.5, $-$0.5) pair (Fig.~\ref{Fig3}; Supplementary Section IX, Fig.~S14). 
Compared to the (3, 0) pair, the magnitude of the Er-Ti coupling is larger overall. 
However, the Rabi frequency does not increase substantially because it is governed by the pair direction as well as the anisotropic nature of $\mathcal{J}$ (Supplementary Section IX).
In addition, our model predicts a Rabi frequency that is approximately 30 times smaller for the Er(O)-Ti pair, owing to the predominantly isotropic exchange coupling (see Supplementary Section XI, Figs.~S19, S20), hence explaining the lack of coherent oscillations in the Er(O)-Ti pair within the range of the experimental control parameters~\cite{reale2024electrically}.

\section{Discussion and outlook}
\noindent
Our demonstration of all-electrical coherent control and readout of both the Er spin and the DQT highlights rare-earth atoms on surfaces as a promising platform for atom-scale spin-based quantum devices.
The atom-by-atom addressability and tunable spin-spin interactions enable a bottom-up design strategy that is difficult to realize in other solid-state qubit architectures.
The high Rabi frequency, reaching up to 190 MHz, allows coherent manipulation on timescales well below the coherence limit, providing access to high-speed quantum gates and strong-drive regimes for exploring nonlinear and higher-order spin dynamics.

Moreover, we identify a general strategy that transduces electrical excitations into an effective coherent rotation of the rare-earth spin.
This mechanism, relying on the anisotropic exchange interaction in coupled spins, not only enables ultra-fast manipulation of rare-earth spins, but also illustrates a broadly transferable design strategy for efficiently driven spin qubits in other scalable platforms.
For example, such a strategy could directly benefit rare-earth ions in solids, which already combine long coherence times with optical interfaces~\cite{zhong2015optically,ranvcic2018coherence,ortu2018simultaneous,le2021twenty,hiraishi2025long}, as well as semiconductor spin qubits~\cite{van2013fast,froning2021ultrafast,wang2022ultrafast,geyer2024anisotropic}, opening new opportunities for high-fidelity, versatile qubit architectures.
Of particular interest, the coherent control mechanism we propose here may also serve as a design concept for molecular spin qubits~\cite{gaita2019molecular,wasielewski2020exploiting,pedersen2016toward} with high and tunable Rabi frequency, where the spin-spin interaction can be tailored through precise engineering of the chemical environment.

\noindent
\section*{Methods}

\subsection{Experiments}
\noindent
All STM measurements were conducted in a home-built low-temperature STM operated at 1.0 K~\cite{hwang2022development}, equipped with a three-axis magnet (1$-$4$-$6 T).
The magnetic tip was made by picking up several Fe atoms from the MgO surface~\cite{Baumann_Paul_science_2015}.
All data were taken at 1.0 K.
Radio-frequency voltages (Keysight E8257D and E8267D) were applied to the junction through the tip.
For time-resolved measurements, the pulse configuration was enabled by an Arbitrary Waveform Generator (Tektronix AWG 5400).
The clocks were synchronized through the data acquisition system (NI USB-6200).
For measurement with multiple RF pulses, the RF voltages were combined through a power splitter (Mini-Circuits ZC2PD-K0244++) and the pulse shapes were controlled by the Arbitrary Waveform Generator.
The surface of a Ag(100) substrate was cleaned by repeated cycles of $\rm Ar^+$ sputtering and annealing at $\sim$700 K. 
Ultrathin MgO(100) films were grown on Ag(100) by evaporating Mg metal in an oxygen atmosphere ($\rm 10^{-6}$ Torr) with the sample kept at a temperature of $\sim$600 K~\cite{Paul_Yang_natphys_2017}. 
Fe, Ti and Er atoms were deposited from high-purity rods ($>$ 99\%) using an e-beam evaporator at cryogenic temperature ($\sim$ 10 K) to have well-isolated atoms on the surface.

\subsection{Modeling}
\noindent
As established in Figs.~2 and 3 of the main text, the effective field acting on the Er spin depends not only on the Er-Ti distance, but also on the direction of the Er-Ti bond. We therefore describe the spin-spin interaction using an anisotropic exchange tensor,
\begin{equation}
    \mathcal{H}_{\rm int}
    =
    \boldsymbol{S}_{\rm Ti}
    \mathcal{J}
    \boldsymbol{S}_{\rm Er},
    \qquad
    \mathcal{J}
    =
    \mathcal{J}
    \left(
    \boldsymbol{q}
    \right).
    \label{eq:H_static_exchange_method}
\end{equation}
Under RF excitation, the modulation of atomic orbitals for both spins can be represented as if the microscopic configuration $\boldsymbol{q}$ were weakly modulated along a trajectory $X_{\rm RF}(t)$. 
This covariant change can be written as the Lie derivative
\begin{equation}
    \delta\mathcal{J}(t)
    =
    \mathcal{L}_{X_{\rm RF}(t)}\mathcal{J},
    \label{eq:Lie_derivative_method}
\end{equation}
with
\begin{equation}
    \mathcal{L}_{X_{\rm RF}(t)}\mathcal{J}
    =
    X_{\rm RF}(t)\cdot\nabla\mathcal{J}
    +
    A_{\rm Ti}^{\mathsf T}(t)\mathcal{J}
    +
    \mathcal{J}A_{\rm Er}(t).
    \label{eq:Lie_decomposition_method}
\end{equation}
Here $A_{\rm Ti}(t)$ and $A_{\rm Er}(t)$ are infinitesimal generators associated with rotations of the Ti and Er local spin frames along the trajectory $X_{\rm RF}(t)$. The first term describes changes of the tensor components due to bond or orbital deformation, whereas the last two terms describe the rotation of the tensor indices. 
The ordinary derivative terms,
\begin{equation}
    X_{\rm RF}(t)\cdot\nabla\mathcal{J}
    =
    \delta\boldsymbol{r}(t)\cdot\nabla_{\boldsymbol r}\mathcal{J}
    +
    \delta\boldsymbol{\eta}_{\rm Ti}(t)\cdot
    \nabla_{\boldsymbol{\eta}_{\rm Ti}}\mathcal{J}
    +
    \delta\boldsymbol{\eta}_{\rm Er}(t)\cdot
    \nabla_{\boldsymbol{\eta}_{\rm Er}}\mathcal{J},
    \label{eq:ordinary_derivative_method}
\end{equation}
do not reproduce the observed angular dependence. We therefore focus on the rotational part of Eq.~\eqref{eq:Lie_decomposition_method}. For a finite angle rotation, this contribution can be expressed as
\begin{equation}
    \mathcal{J}
    \rightarrow
    \mathcal{J}_{\rm rot}(t)
    =
    R_{\rm Ti}^{\mathsf T}(t)
    \mathcal{J}
    R_{\rm Er}(t),
    \label{eq:J_rot_method}
\end{equation}
and hence
\begin{equation}
    \delta\mathcal{J}_{\rm rot}(t)
    =
    R_{\rm Ti}^{\mathsf T}(t)
    \mathcal{J}
    R_{\rm Er}(t)
    -
    \mathcal{J}.
    \label{eq:deltaJ_rot_method}
\end{equation}
The corresponding driving Hamiltonian under RF excitation is
\begin{equation}
    \mathcal{H}_{1}(t) =
    \boldsymbol{S}_{\rm Ti}
    \delta\mathcal{J}_{\rm rot}(t)
    \boldsymbol{S}_{\rm Er},
    \label{eq:H_rot_method}
\end{equation}
which is a more general form of Eq.~\eqref{EQ7}.
A more detailed description of the theory is provided in Supplementary  Section IX.

\section*{Acknowledgment}
\noindent
The authors acknowledge support from Institute for Basic Science (grant: IBS-R027-D1).
We thank Harald Brune, Nicol\'as Lorente, Chirstopher P. Lutz and Christoph Wolf for the stimulating discussions.

\section*{Author Contribution}
\noindent
FD conceived the experiment. YL, DC, SR, JO, SP, and FD performed the experiments.
YL and FD performed the modeling.
FD and SC performed the multiplet calculations.
YL, AJH, SP and FD wrote the manuscript with the input from all authors.

\clearpage

\bibliography{references}

\clearpage

\clearpage

\clearpage

\clearpage

\clearpage

\clearpage


\setcounter{page}{1}
\setcounter{section}{0}
\setcounter{figure}{0}
\setcounter{table}{0}
\setcounter{equation}{0}
\renewcommand{\thefigure}{S\arabic{figure}}
\renewcommand{\thetable}{S\arabic{table}}
\renewcommand{\theequation}{S\arabic{equation}}
\renewcommand*{\thepage}{S\arabic{page}}
\onecolumngrid
\begin{center}
{\large \textbf{Supplementary Information for \\ ``\ourtitle"}}\\
\end{center}


\section{Basic characterization of Er(B)}
\subsection{Topography and differential conductance}

\begin{figure}[htbp]
    \centering
    \includegraphics[width = 0.7\textwidth]{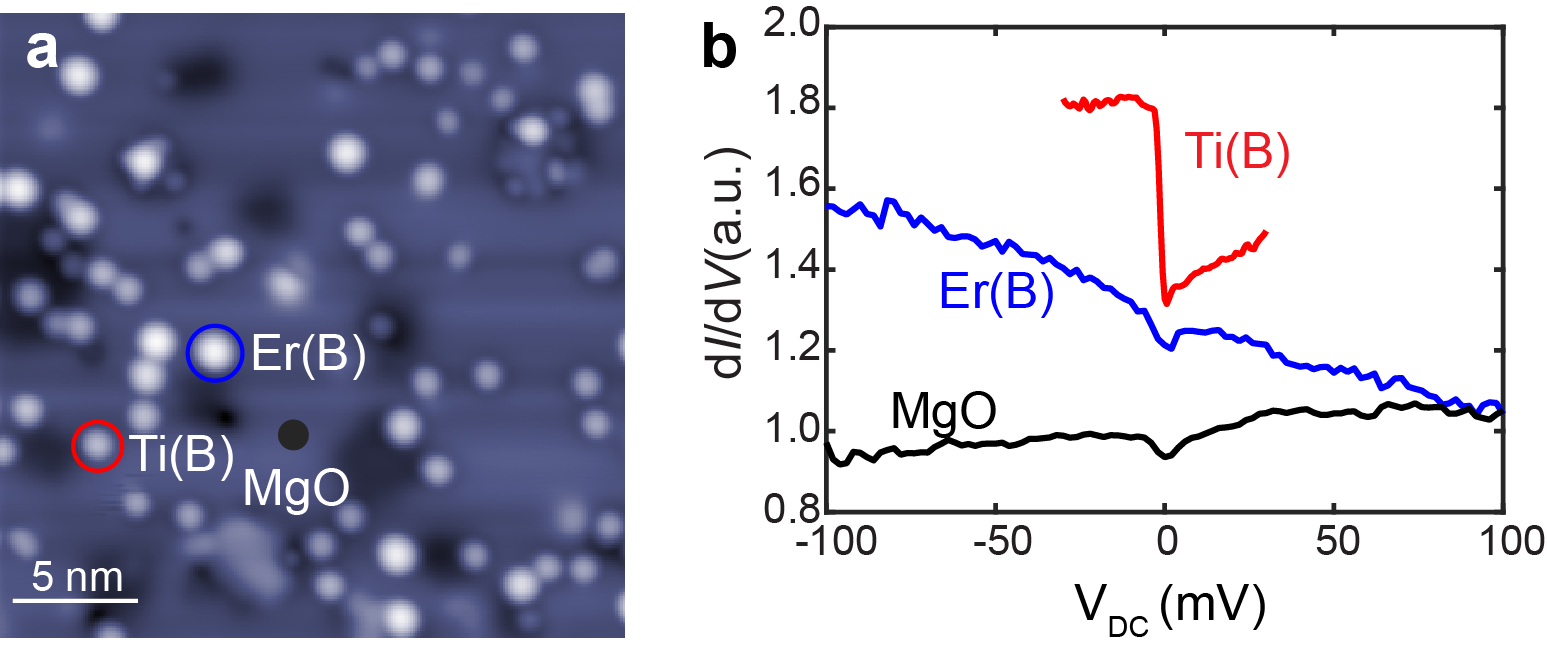}
    \caption{\textbf{Topography and differential conductance of Er(B) and Ti(B).} \textbf{a,} STM topographic image of a MgO patch with Er, Ti and Fe atoms ($V_{\rm DC}$ = 100 mV, $I_{\rm DC}$ = 20 pA, 20 nm $\times$ 20 nm). An Er(B) and a Ti(B) atoms are marked by the blue and red circles, respectively. \textbf{b,} Differential conductance spectra using a spin-polarized tip on Er(B), Ti(B) and MgO in (\textbf{a}) ($V_{\rm DC}$ = 100 mV (blue and black curves) and 30 mV (red curve), $I_{\rm DC}$ = 200 pA, $V_{\rm osc}$ = 1 mV, $\boldsymbol{B}_{\rm ext}$ = 0.25 T, $\theta$ = 90$\degree$ and $\phi$ = 52$\degree$).}
    \label{FigS_spectra}
\end{figure}

\noindent
Er(B) appears higher than both Er(O) and Ti(B) in STM topographic images acquired at
$V_{\rm DC}=100~{\rm mV}$ and $I_{\rm DC}=20~{\rm pA}$, with an apparent height of
$265~{\rm pm}$, as highlighted by the blue circle in
Fig.~\ref{FigS_spectra}\textbf{a}. The differential conductance
(${\rm d}I/{\rm d}V$) spectra measured on Ti(B) and Er(B) with a spin-polarized
tip are shown in Fig.~\ref{FigS_spectra}\textbf{b}. Ti(B) exhibits a clear
spin-polarized signature, manifested as a step-like feature around zero bias.
By contrast, the ${\rm d}I/{\rm d}V$ spectra measured on Er(B) show no
observable spin excitation or spin-polarized contrast~\cite{reale2024electrically},
consistent with the closed-shell character of its outer electronic states.

\subsection{ESR on Er(B)}
\begin{figure}[htbp]
    \centering
    \includegraphics[width = 0.6\textwidth]{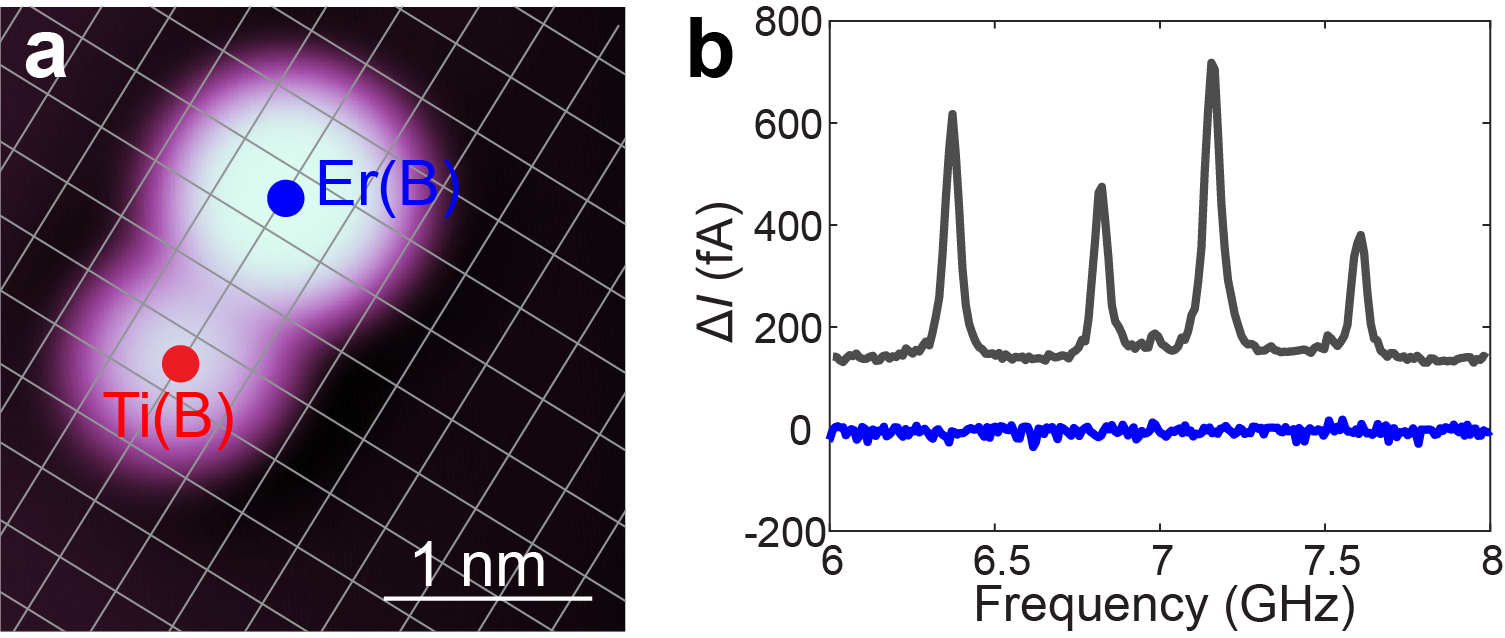}
    \caption{\textbf{ESR spectra on Er(B) and Ti(B).} \textbf{a,} STM topography of an Er-Ti pair ($V_{\rm DC}$ = 100 mV, $I_{\rm DC}$ = 20 pA, 3 nm$\times$ 3 nm). The gray grid represents the MgO lattice with the crossing points representing oxygen atomic sites on the surface layer. \textbf{b,} ESR spectra at the two spots in (\textbf{a}) ($V_{\rm DC}$ = 100 mV, $I_{\rm DC}$ = 20 pA, $V_{\rm RF}$ = 20 mV, $\boldsymbol{B}_{\rm ext}$ = 0.25 T, $\theta$ = 90$\degree$ and $\phi$ = 52$\degree$).}
    \label{FigS1}
\end{figure}

\noindent
Here we provide ESR spectra on both atoms in an Er-Ti pair.
Figure~\ref{FigS1} shows a featureless spectrum when we place the tip on Er(B), which is due to the closed outer electron shell of an Er atom on an ultrathin MgO surface.

\section{Transition energies of $f_5$}
\begin{figure}[htbp]
    \centering
    \includegraphics[width = 0.4\textwidth]{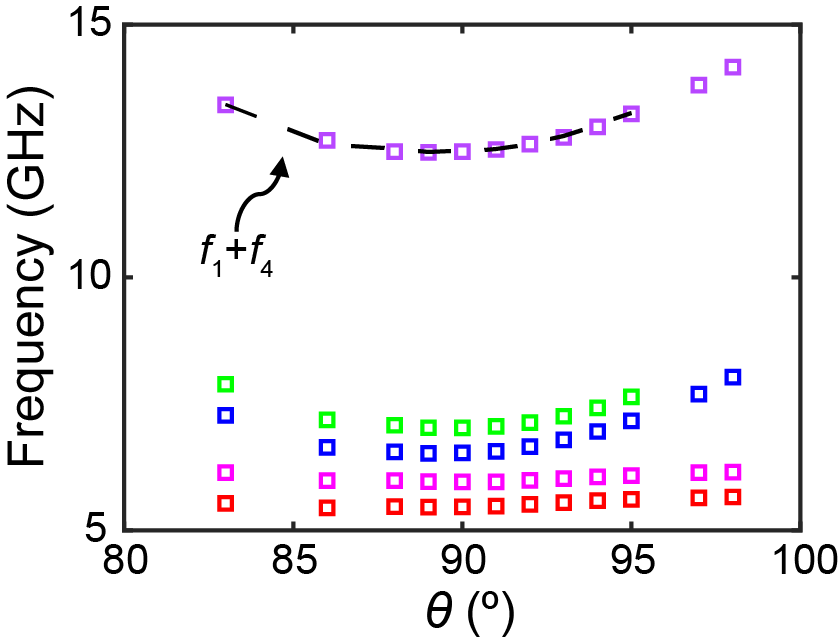}
    \caption{\textbf{Double-quantum transition $f_5$.} ESR peak position for each transition shown in Fig.~1\textbf{c} in the manuscript. }
    \label{FigS2}
\end{figure}

\noindent
As shown in Fig.~\ref{FigS2}, the transition energy of $f_5$ remains locked to the two-transition sum, $f_1+f_4$ or equivalently $f_2+f_3$, over the entire polar-angle $\theta$ sweep.

\clearpage

\section{Fitting ESR spectrum}
\noindent
In this section, we describe the fitting procedure used to analyze the ESR spectra presented in Fig.~4 of the main text, from which we extract the effective exchange interaction and the anisotropic $\mathfrak{g}$-tensor. To determine the effective $g$ factors of Er on the two-monolayer MgO surface, we map the ground-state doublet of Er to an effective spin-$1/2$ degree of freedom, denoted by $\boldsymbol{S}_{\rm Er}$. Within this effective-spin description, the Er-Ti pair is described by a spin Hamiltonian consisting of the Zeeman term $\mathcal{H}_{\rm Z}$ and the spin-spin interaction term $\mathcal{H}_{\rm int}$:
\begin{equation}
    \label{eq.1}
   \mathcal{H} = \mathcal{H}_{\rm Z}+\mathcal{H}_{\rm int}= -\mu_{\rm B}\boldsymbol{S}_{\rm Ti} \cdot \mathfrak{g}^{\rm Ti}\cdot(\boldsymbol{B}_{\rm ext}+\boldsymbol{B}_{\rm tip}) -\mu_{\rm B}\boldsymbol{S}_{\rm Er} \cdot \mathfrak{g}^{\rm Er}\cdot\boldsymbol{B}_{\rm ext} + \boldsymbol{S}_{\rm Ti} \cdot \mathcal{J}\cdot \boldsymbol{S}_{\rm Er},
\end{equation}
\noindent with $\mu_{\rm B}$ the Bohr magneton, $\boldsymbol{B}_{\rm ext}$ the external magnetic field, $\boldsymbol{B}_{\rm tip}$ the magnetic field from the tip, $\mathfrak{g}^{\rm Ti(Er)}$ the $\mathfrak{g}$-tensor for Ti (Er) spin and $\mathcal{J}$ the spin interaction tensor. As in previous works, the effective tip field $\boldsymbol{B}_{\rm tip}$ only applies to the spin under the tip, in this case Ti~\cite{yang2017engineering}.
The $\mathcal{H}_{\rm int}$ term can be decomposed into exchange and dipole-dipole contributions:
\begin{equation}
    \label{eq.3}
    \mathcal{H}_{\rm exc} =\boldsymbol{S}_{\rm Ti} \cdot \mathcal{J}^{\rm exc}\cdot \boldsymbol{S}_{\rm Er},
\end{equation}
\begin{equation}
\label{eq.2}
\mathcal{H}_{\rm dip} = \frac{\mu_0\mu_B^2}{4\pi r^3}[(\mathfrak{g}^{\rm Ti}\cdot\boldsymbol{S}_{\rm Ti} )\cdot(\mathfrak{g}^{\rm Er}\cdot\boldsymbol{S}_{\rm Er} )-3(\hat{r}\cdot \mathfrak{g}^{\rm Ti}\cdot\boldsymbol{S}_{\rm Ti})(\hat{r}\cdot \mathfrak{g}^{\rm Er}\cdot\boldsymbol{S}_{\rm Er})].
\end{equation}
\noindent Here $\hat{r}$ is the unit vector connecting the two spins and $\mathcal{J}^{\rm exc}$ is the exchange tensor.
In this model, we assume that the tip-induced magnetic field,
$\boldsymbol{B}_{\rm tip}$, follows the direction of the external magnetic field.
The experimental data are fitted using this Hamiltonian, with both the transition
energies and the normalized Rabi frequencies included in the fitting procedure.
Here, the Rabi frequencies are normalized by the Rabi frequency of the $f_3$
transition at $\theta=90\degree$ and $\phi=52\degree$, as shown in
Fig.~\ref{FigS_wobble}\textbf{c}. Including the Rabi frequencies provides direct
constraints on the driving matrix elements, thereby enabling a more stringent and
physically faithful characterization of the microscopic driving mechanism.

Using this spin Hamiltonian, we fit the angular dependence of ESR transitions for the (3, 0) pair presented in Fig.~4 of the main text, and we obtain the  Er $\mathfrak{g}$-tensor:
\begin{equation}
    \mathfrak{g}^{\rm Er}=\begin{pmatrix}
        \mathfrak{g}^{\rm Er}_{xx}&0&0\\
        0&\mathfrak{g}^{\rm Er}_{yy}&0\\
        0&0&\mathfrak{g}^{\rm Er}_{zz}
    \end{pmatrix}
\end{equation}
with $\mathfrak{g}^{\rm Er}_{xx}$ = 1.05 $\pm$ 0.02, $\mathfrak{g}^{\rm Er}_{yy}$ = 2.22 $\pm$ 0.02, and $\mathfrak{g}^{\rm Er}_{zz}$ = 9.59 $\pm$ 0.04.
The fitted $\mathcal{J}^{\rm exc}$ from the fitting in MHz is:
\begin{equation}
    \label{EQ.S3}
    \mathcal{J}^{\rm exc}=\begin{pmatrix}
\mathcal{J}^{\rm exc}_{xx} & 0 & 0\\
0 & \mathcal{J}^{\rm exc}_{yy} & 0\\
0 & 0   & \mathcal{J}^{\rm exc}_{zz}
\end{pmatrix}
\end{equation}
with $\mathcal{J}^{\rm exc}_{xx}$ = 79.6 $\pm$ 5.5 MHz, $\mathcal{J}^{\rm exc}_{yy}$ = 543.7 $\pm$ 6.3 MHz, and $\mathcal{J}^{\rm exc}_{zz}$ = 1272.5 $\pm$ 7.8 MHz.

\section{Decomposition of $\ket{+}$ state}
\noindent
In this section, we examine how the spin character of the $f_3$ transition evolves with the polar angle $\theta$, while keeping the azimuthal angle fixed at $\phi=52\degree$. We first calculate the eigenstates of the effective spin-$1/2$ Hamiltonian in Eq.~\eqref{eq.1}, including the full spin-spin interaction with fitted $\mathcal{J}$ and $\mathfrak{g}^{\rm Er}$ tensors, and label the four resulting energy eigenstates as $\ket{00}$, $\ket{-}$, $\ket{+}$, and $\ket{11}$, following the notation used in the main text. We then calculate the eigenstates of the corresponding non-interacting Hamiltonian, for which the eigenstates are pure Zeeman product states denoted by $\ket{00}$, $\ket{10}$, $\ket{01}$, and $\ket{11}$.

\begin{figure}[htbp]
    \centering
    \includegraphics[width = 0.5\textwidth]{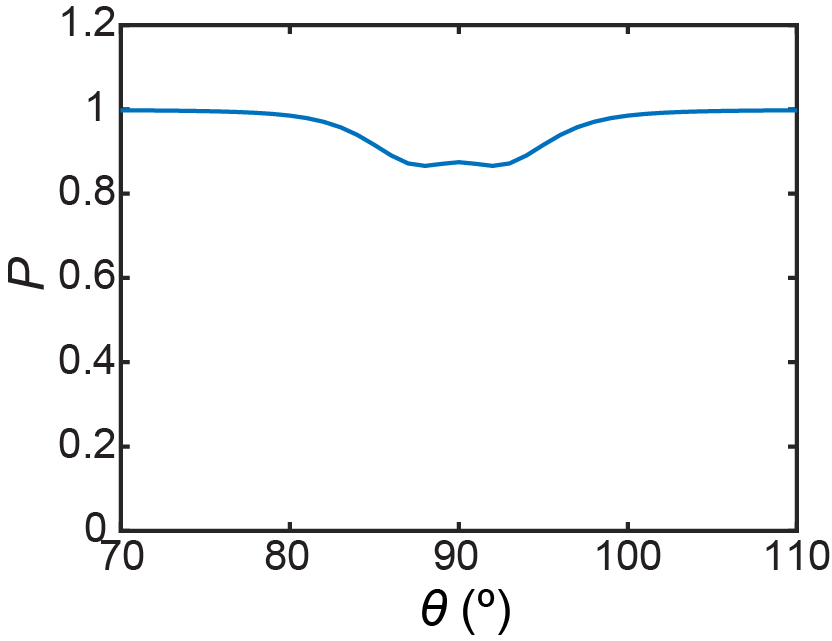}
    \caption{\textbf{Decomposition of $\ket{+}$ state.} $P=|{\langle +| 01\rangle }|^2$ as a function of polar angle $\theta$ with azimuthal angle fixed at 52$\degree$, $\boldsymbol{B}_{\rm ext}$ = 0.25 T.}
    \label{FigS_mixing}
\end{figure}

To quantify the Er contribution to the $f_3$ transition, we evaluate the overlap
\begin{equation}
    P=\left|\braket{+|01}\right|^2,
\end{equation}
as shown in Fig.~\ref{FigS_mixing}. At $\theta=83\degree$ and $\phi=52\degree$, which is the field condition for the Rabi data presented in Fig.~2 of the main text, we find that the $\ket{+}$ state has predominantly Er character, containing $96\%$ Er-like character and only $4\%$ Ti-like character.

Note that the Rabi rate map for SQT Er in Fig.~5 of the main text is obtained at each field direction by:
\begin{equation}
    \label{EQ_SQT_ER}
    \Omega=\bra{00}\mathcal{H}_{\rm drive}\ket{01}
\end{equation}
where $\ket{01}$ is the pure Zeeman energy state described above and $\mathcal{H}_{\rm drive}$ is the driving Hamiltonian, which will be discussed in the following sections.

\section{Influence of tip-induced field}
\begin{figure}[htbp]
    \centering
    \includegraphics[width = 0.5\textwidth]{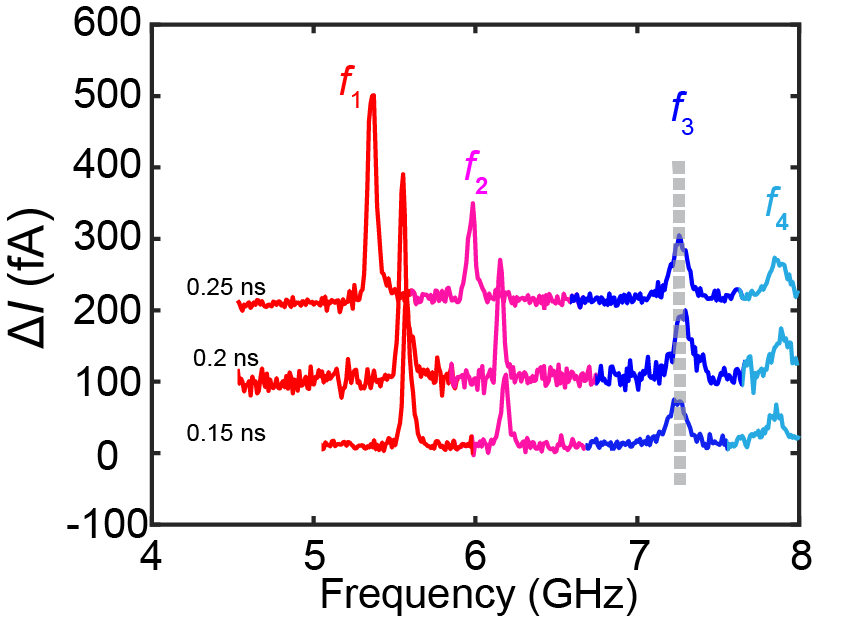}
    \caption{\textbf{Influence of tip-induced field.} ESR spectra taken under the field condition of $\boldsymbol{B}_{\rm ext}$ = 0.25 T, $\theta$ = 83$\degree$ and $\phi$ = 52$\degree$ as a function of tip-atom distance parameterized by the conductance. All data were taken using an RF voltage of $V_{\rm RF}$ = 20 mV.}
    \label{FigS_Tip}
\end{figure}

\noindent
In Fig.~\ref{FigS_Tip}, we show that the tip-induced magnetic field has a pronounced influence on the transition energies of $f_1$ and $f_2$, while its effect on $f_3$ and $f_4$ is negligible under the field condition used in Figs.~1 and 2 of the main text
($\boldsymbol{B}_{\rm ext}=0.25~{\rm T}$, $\theta=83\degree$, and $\phi=52\degree$).
This suggests that $f_1$ and $f_2$ are predominantly Ti transitions while $f_3$ and $f_4$ are predominantly Er transitions.

\clearpage

\section{Dependence of Rabi frequencies of $f_1$ and $f_3$ on tip-atom distance}
\begin{figure}[htbp]
    \centering
    \includegraphics[width = 0.75\textwidth]{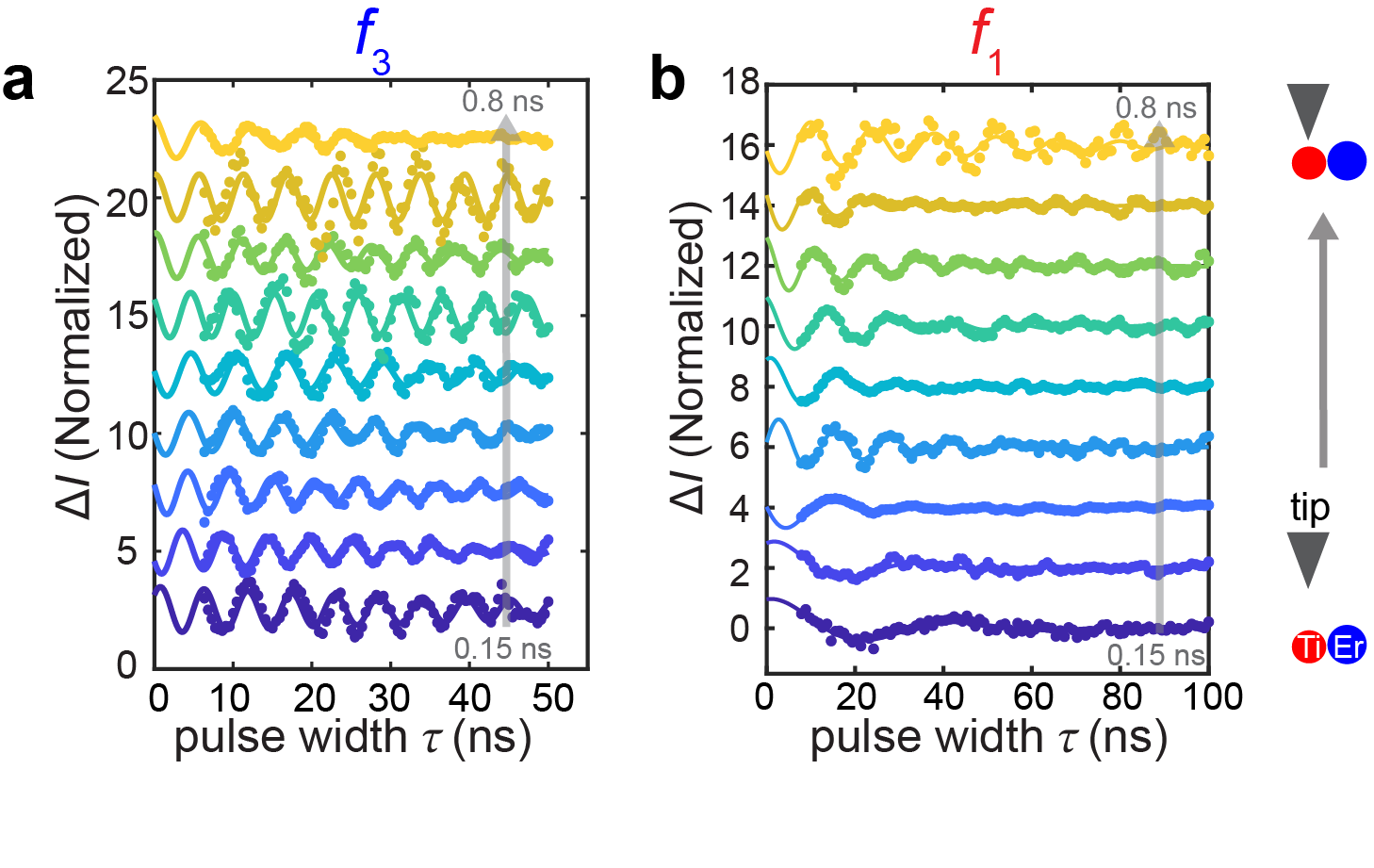}
    \caption{\textbf{Rabi oscillations of $f_1$ and $f_3$ with different tip-Ti atom distances. }  \textbf{a,} Rabi oscillations for $f_3$ with different conductances.  \textbf{b,} Rabi oscillation for $f_1$ with different conductance. All data were taken with $V_{\rm RF}$ = 100 mV, $\boldsymbol{B}_{\rm ext}$ = 0.25 T, $\theta$ = 83$\degree$, $\phi$ = $52\degree$.}
    \label{FigS_Ti_conduc}
\end{figure}

\noindent
To investigate the influence of the tip on the Rabi frequency of the Ti-like transition $f_1$ and the Er-like transition $f_3$, we performed Rabi measurements on $f_1$ and $f_3$ with different tip-Ti distances ($V_{\rm RF}$ = 100 mV, $\boldsymbol{B}_{\rm ext}$ = 0.25 T, $\theta$ = 83$\degree$, $\phi$ = $52\degree$).
As shown in Fig.~\ref{FigS_Ti_conduc}, when the tip approaches closer to the Ti atom, the Rabi frequency first increases and then decreases. In contrast, for the $f_3$ transition, the Rabi oscillation frequency does not exhibit a clear correlation with the tip-Ti atom distance, consistent with the discussion in the main text. 
The extracted Rabi frequency as a function of conductance is shown in Fig.~2(\textbf{e}) of the main text. 

As the tip initially approaches the Ti atom, the enhanced exchange interaction between the spin-polarized tip and the Ti spin increases the effective driving field~\cite{phark2023electric}, leading to an enhancement of the Rabi frequency of the $f_1$ transition. In this regime, the dominant effect is the strengthening of the transverse driving component mediated by the tip polarization.
However, when the tip is brought even closer, the tip-induced exchange field becomes sufficiently strong to modify the quantization axis of the Ti spin. This additional static polarization effect alters the spin eigenbasis and can consequently change the effective transverse driving amplitude. As a result, the Rabi frequency no longer follows a simple monotonic trend with conductance. 

\section{Spin dynamics of Er $4f$ spin}
\begin{figure}[htbp]
    \centering
    \includegraphics[width = 1\textwidth]{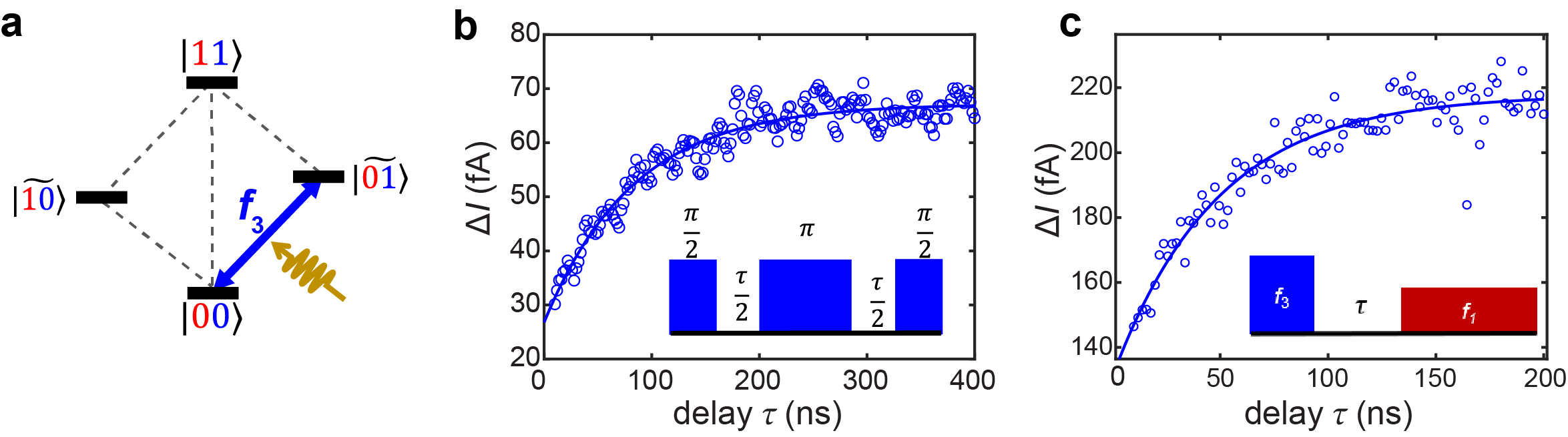}
    \caption{\textbf{Er spin dynamics.}
\textbf{a,} Energy diagram of Er-Ti.
\textbf{b,} Spin-echo measurement performed on $f_3$ ($V_{\rm DC} = 100$ mV, $I_{\rm DC} = 20$ pA, $V_{\rm RF} = 30$ mV).
\textbf{c,} Recovery-sequence measurement performed on $f_3$ ($V_{\rm DC} = 100$ mV, $I_{\rm DC} = 20$ pA, $V_{\rm RF} = 30$ mV for pumping $f_3$ and $V_{\rm RF} = 20$ mV for probing).
All data were acquired under a magnetic-field condition of $\boldsymbol{B}_{\rm ext} = 0.25$ T, $\theta = 90\degree$, and $\phi = 52\degree$ ($f_{\rm res}$ = 6.51 GHz).} 
    \label{FigS_dyn}
\end{figure}

\noindent
In this section, we present the spin dynamics of the Er $4f$  spin. We performed a standard spin-echo measurement on the $f_3$ transition under the magnetic-field condition ($\boldsymbol{B}_{\rm ext} = 0.25$ T, $\theta = 90\degree$, $\phi = 52\degree$). From this measurement, we extracted an echo coherence time of $T_2^{\rm Echo} = 112.9 \pm 13.4$ ns, as shown in Fig.~\ref{FigS_dyn}\textbf{b}.
Using a recovery pulse sequence, we further measured the spin lifetime of the $f_3$ transition, obtaining $T_1 = 57.2 \pm 9.8$ ns, as presented in Fig.~\ref{FigS_dyn}\textbf{c}.
We attribute the relatively short coherence time and lifetime primarily to the presence of Ti as the coupling partner of Er.
Replacing Ti with a different neighboring atom, such as Ho or Sm, may mitigate these limitations. Moreover, implementing an Er-Ho or Er-Sm pair as a remote qubit, read out via a Ti sensor spin, could substantially enhance the coherence properties of the system.

\clearpage

\section{Multiplet calculations}
\noindent
Quantum states of Er atoms were computed using the Quanty multielectron code~\cite{haverkort2016quanty}. Open-shell multiplet calculations were performed by modeling the intra- and inter-orbital electron-electron interactions using Slater integrals, whose values were obtained from the full-electron Cowan atomic structure code~\cite{cowan2023theory}. All Slater integrals were calculated assuming a $4f^{11}5d^2$ electronic configuration. The two electrons in the $5d$ shell were included to account for configuration interaction mediated by the ligand field, as discussed below.
The Slater integrals were rescaled to account for screening effects arising from surface electrons. A reduction factor of 0.85 was adopted based on previous X-ray measurements~\cite{reale2023erbium}. In addition to electron-electron interactions, the Hamiltonian used for the multiplet calculations includes spin-orbit coupling, Zeeman energy due to the external magnetic field, crystal field effects acting on the different shells, and configuration-interaction ligand field terms acting on the outer $5d$ electrons. The spin-orbit coupling parameters were also computed using Cowan’s atomic structure code~\cite{reale2023erbium}. The crystal field acting on the open valence shells of the lanthanide was modeled according to the symmetry and orbital character of the corresponding states.

For localized $4f$ electrons, we modeled the crystal field using a point-charge electrostatic model~\cite{gorller1996rationalization,uldry2012systematic}, including only the nearest neighbors of the adsorbed lanthanide atom. The positions of the neighboring ions surrounding the Er atom were obtained from DFT calculations~\cite{reale2023erbium}, and the Born charges ($\pm 2e$ for Mg and O, respectively) were globally rescaled to account for mutual screening and the finite spatial extension of the ions.
In addition, the radial extension of the $4f$ orbitals was renormalized by rescaling the radial operator as $\hat{r} \rightarrow \alpha\hat{r}$ to account for dielectric screening by surface electrons. The reduced charge ($q_{red}$) and the radial rescaling parameter ($\alpha$) were treated as free parameters to fit the experimental $\mathfrak{g}$-tensor. To this end, the three principal values of the $\mathfrak{g}$-tensor were calculated from the Zeeman energy along the three principal axes by projecting onto a pseudo-spin $1/2$ ground-state doublet. However, with only these two degrees of freedom, it was not possible to reproduce the experimentally measured $\mathfrak{g}$-tensor with satisfactory accuracy.

To address this limitation, we introduced a finite electron occupation of the $5d$ orbitals, which provides an additional anisotropic interaction acting on the $4f$ electrons. For the $5d$ shell, an artificially large crystal-field splitting (10 eV) was imposed to separate the $5d_{y^2}$ type orbital from the remaining $5d$ orbitals. This choice isolates the orbital with maximal overlap toward the positively charged Mg atoms, which is expected to lie lowest in energy. Previous DFT studies of lanthanide atoms on MgO/Ag(100)~\cite{singha2021mapping,donati2021correlation} report a typical $5d$ occupation below one electron and no net polarization within this shell.

To reproduce this situation, we included a ligand-field term with configuration interaction between the $5d^0$ and $5d^2$ configurations. The exclusion of odd $5d$ configurations ensures negligible net polarization. The $5d$ occupation was tuned by fixing the hopping parameter to 0.5 eV and treating the on-site energy $\Delta$ as an additional free parameter in the fit.
The fitting procedure was performed using Bayesian optimization as implemented in MATLAB. The error function was defined as the squared difference between calculated and experimental principal values of the Er $\mathfrak{g}$-tensor. The parameters reported in Fig.~4 of the main text correspond to the best-fit values $\alpha = 1.155$, $q_{\rm red} = 0.5932$, and $\Delta = 6.6467$~eV, yielding a $5d$ occupation of 0.5344 electrons.

\begin{figure}[htbp]
    \centering
    \includegraphics[width = 0.5\textwidth]{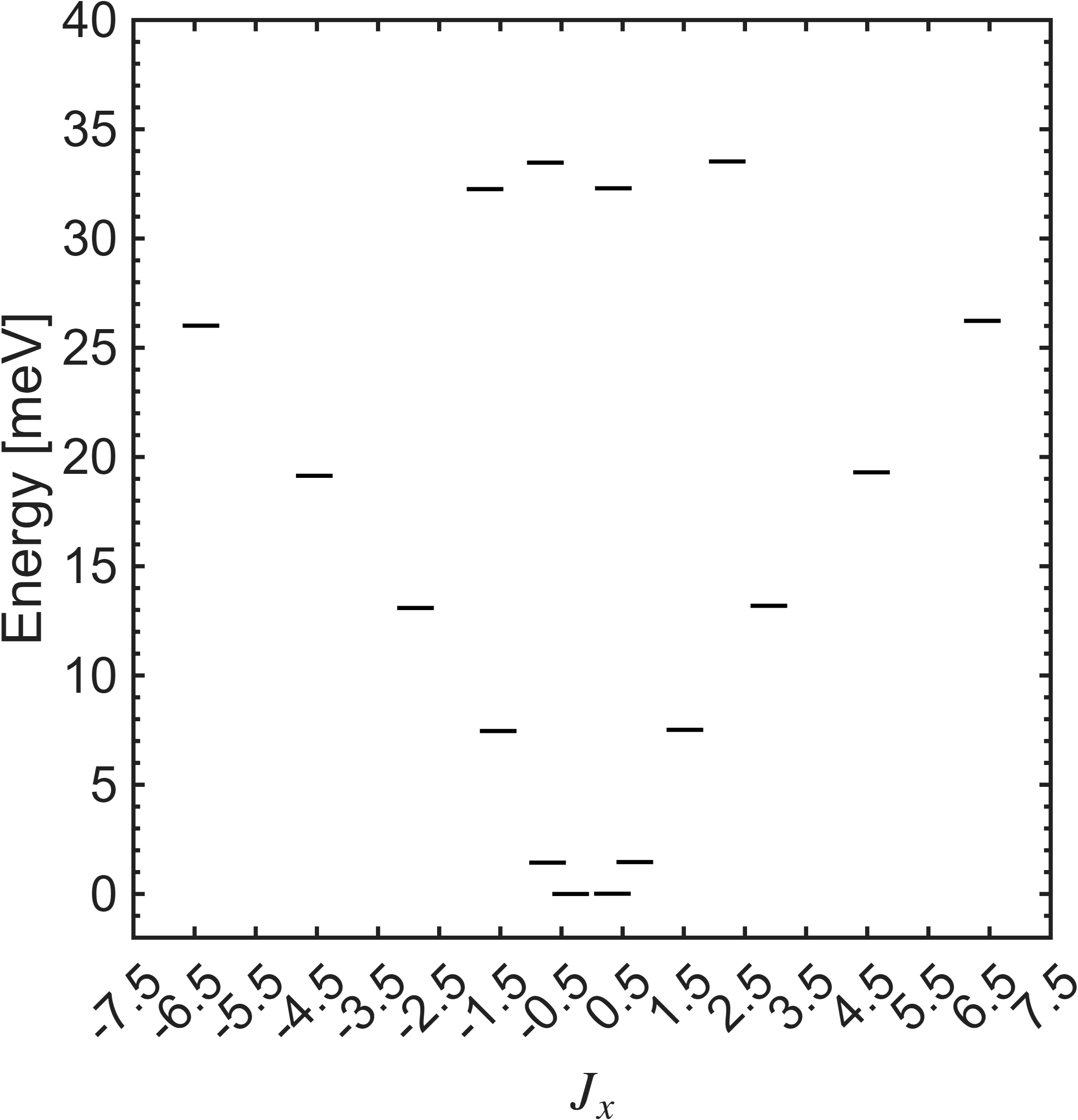}
    \caption{\textbf{Quantum levels of Er(B)}. Eigenstates of the lowest  $^4I_{15/2}$ multiplet of Er(B) versus their total angular moment projected along the oxygen $x$ direction ($J_x$).}
    \label{FigS_Er_Jx}
\end{figure}

The eigenstates of the lowest $^{4}I_{15/2}$ multiplet split due to the combination of crystal field from the point charge and electrostatic interaction with the $5d$ electron into a set of doublets, whose angular momentum distribution is best visualized by projecting along the oxygen direction ($x$). Along this axis (Fig.~\ref{FigS_Er_Jx}), the eigenstates form an upturned parabola representative of a hard axis. Similar to Er(O), the lowest two states are minimally projected along the quantization axis, although the two atoms have different hard axis directions, namely, out-of-plane for Er(O)~\cite{reale2023erbium, reale2024electrically} and in-plane along O-O direction for Er(B). 
The hard-axis anisotropy along the oxygen direction also appears in the $\mathfrak{g}$-tensor shown in Fig.~4\textbf{c} of the main text, where the $x$-component is the lowest among the three principal directions.

Minimally projected states in Kramers ions typically provide a natural two-level system to enable spin transitions through exchange of a single unit of angular momentum~\cite{reale2023erbium, reale2024electrically}, which is the key to efficient quantum coherent control. To verify whether this holds for Er(B) multiplet states, we expand the eigenstates $\ket{i}$ from the full Hamiltonian over the free atom basis $\ket{J = 15/2, m_J}$ obtained by taking the quantization axis along the oxygen direction:
\begin{equation}
    \ket{i} = \sum_{j} c_{ij} \ket{J=15/2, m_J}.
    \label{eq:free_atom_projection}
\end{equation}
The square moduli of the coefficients ($\left|c_{ij}\right|^2)$ representing the relative weight of each $\ket{J=15/2, m_J}$ in the Er(B) eigenstates are shown in Table~\ref{tab:cf_eigenstates}. In line with our expectation, the lowest doublet shows the largest weights on the $\ket{J=15/2, m_J = \pm 1/2}$ states, which enables efficient first-order ESR transitions. In addition, this decomposition further justifies modeling the ESR data with a spin Hamiltonian where the lowest two levels are mapped into an effective S = 1/2 system.  

\begin{table*}[t]
\centering
\caption{Eigenstates of Er(B) obtained using multiplet calculations. The left section shows the expectation values of the projected angular momentum along the oxygen direction $J_x$, the related magnetic moment $\mu_x$, and the energy $E$ in meV with respect to the ground state. The right section shows the relative weight $\left|c_{ij}\right|^2$ of the crystal-field eigenstates $\ket{i}$ expressed in the free atom $\ket{J = 15/2, m_J}$ basis, and calculated for a quantization axis along the oxygen direction with a magnetic field $\boldsymbol{B} = 0.25$ T along the $x$-axis ($\theta = 90 \degree$ and $\phi = 0 \degree$).}
\label{tab:cf_eigenstates}
\scriptsize
\setlength{\tabcolsep}{3.0pt}
\renewcommand{\arraystretch}{1.12}

\resizebox{\textwidth}{!}{%
\begin{tabular}{|c|r r r|rrrrrrrrrrrrrrrr|}
\hline
State & $J_x$ & $\mu_x$ & $E$ (meV)
& \multicolumn{16}{c|}{$m_J$ component weight ($\left|c_{ij}\right|^2$)} \\
\cline{5-20}
& & &
& $-\frac{15}{2}$ 
& $-\frac{13}{2}$ 
& $-\frac{11}{2}$ 
& $-\frac{9}{2}$ 
& $-\frac{7}{2}$ 
& $-\frac{5}{2}$ 
& $-\frac{3}{2}$ 
& $-\frac{1}{2}$ 
& $\frac{1}{2}$ 
& $\frac{3}{2}$ 
& $\frac{5}{2}$ 
& $\frac{7}{2}$ 
& $\frac{9}{2}$ 
& $\frac{11}{2}$ 
& $\frac{13}{2}$ 
& $\frac{15}{2}$ \\
\hline
1  & $-0.350$ & $ 0.419$ & $ 0.000$
& 0.0 & 0.0 & 0.0 & 0.0 & 0.0 & 10.1 & 0.0 & 79.8
& 0.0 & 2.6 & 0.0 & 7.0 & 0.0 & 0.4 & 0.0 & 0.0 \\
\hline
2  & $ 0.333$ & $-0.399$ & $ 0.012$
& 0.0 & 0.0 & 0.5 & 0.0 & 7.2 & 0.0 & 2.8 & 0.0
& 79.4 & 0.0 & 10.1 & 0.0 & 0.0 & 0.0 & 0.0 & 0.0 \\
\hline
3  & $-0.728$ & $ 0.872$ & $ 1.435$
& 0.0 & 0.0 & 0.0 & 0.0 & 5.1 & 0.0 & 68.8 & 0.0
& 11.6 & 0.0 & 11.4 & 0.0 & 3.0 & 0.0 & 0.0 & 0.0 \\
\hline
4  & $ 0.699$ & $-0.837$ & $ 1.460$
& 0.0 & 0.0 & 0.0 & 3.2 & 0.0 & 12.1 & 0.0 & 11.2
& 0.0 & 68.5 & 0.0 & 5.0 & 0.0 & 0.0 & 0.0 & 0.0 \\
\hline
5  & $-1.535$ & $ 1.840$ & $ 7.459$
& 0.0 & 1.8 & 0.0 & 22.6 & 0.0 & 48.0 & 0.0 & 0.3
& 0.0 & 9.9 & 0.0 & 14.8 & 0.0 & 2.5 & 0.0 & 0.0 \\
\hline
6  & $ 1.518$ & $-1.821$ & $ 7.512$
& 0.0 & 0.0 & 2.7 & 0.0 & 15.2 & 0.0 & 9.1 & 0.0
& 0.3 & 0.0 & 48.7 & 0.0 & 22.3 & 0.0 & 1.8 & 0.0 \\
\hline
7  & $-2.887$ & $ 3.461$ & $13.090$
& 1.1 & 0.0 & 35.6 & 0.0 & 36.0 & 0.0 & 11.9 & 0.0
& 3.4 & 0.0 & 0.8 & 0.0 & 8.6 & 0.0 & 2.7 & 0.0 \\
\hline
8  & $ 2.891$ & $-3.465$ & $13.190$
& 0.0 & 2.9 & 0.0 & 8.5 & 0.0 & 0.5 & 0.0 & 3.2
& 0.0 & 11.5 & 0.0 & 36.8 & 0.0 & 35.4 & 0.0 & 1.1 \\
\hline
9  & $-4.542$ & $ 5.441$ & $19.140$
& 0.0 & 53.3 & 0.0 & 19.1 & 0.0 & 18.4 & 0.0 & 3.0
& 0.0 & 2.0 & 0.0 & 0.1 & 0.0 & 3.6 & 0.0 & 0.5 \\
\hline
10 & $ 4.570$ & $-5.474$ & $19.298$
& 0.5 & 0.0 & 3.5 & 0.0 & 0.0 & 0.0 & 1.9 & 0.0
& 2.9 & 0.0 & 18.3 & 0.0 & 19.7 & 0.0 & 53.3 & 0.0 \\
\hline
11 & $-6.396$ & $ 7.654$ & $26.011$
& 71.7 & 0.0 & 10.5 & 0.0 & 13.7 & 0.0 & 2.2 & 0.0
& 0.5 & 0.0 & 0.2 & 0.0 & 0.1 & 0.0 & 1.1 & 0.0 \\
\hline
12 & $ 6.382$ & $-7.638$ & $26.232$
& 0.0 & 1.1 & 0.0 & 0.2 & 0.0 & 0.1 & 0.0 & 0.5
& 0.0 & 2.2 & 0.0 & 13.9 & 0.0 & 11.0 & 0.0 & 71.1 \\
\hline
13 & $-1.750$ & $ 2.094$ & $32.262$
& 0.0 & 30.2 & 0.0 & 32.3 & 0.0 & 6.2 & 0.0 & 0.3
& 0.0 & 0.1 & 0.0 & 5.1 & 0.0 & 14.6 & 0.0 & 11.2 \\
\hline
14 & $ 0.348$ & $-0.417$ & $32.299$
& 14.6 & 0.0 & 21.2 & 0.0 & 8.3 & 0.0 & 0.4 & 0.0
& 0.1 & 0.0 & 4.5 & 0.0 & 25.8 & 0.0 & 25.0 & 0.0 \\
\hline
15 & $-0.764$ & $ 0.914$ & $33.474$
& 12.1 & 0.0 & 26.0 & 0.0 & 14.5 & 0.0 & 3.0 & 0.0
& 1.8 & 0.0 & 6.1 & 0.0 & 20.5 & 0.0 & 16.0 & 0.0 \\
\hline
16 & $ 2.211$ & $-2.644$ & $33.526$
& 0.0 & 10.5 & 0.0 & 14.2 & 0.0 & 4.5 & 0.0 & 1.6
& 0.0 & 3.2 & 0.0 & 17.3 & 0.0 & 32.5 & 0.0 & 16.1 \\
\hline
\end{tabular}
}
\end{table*}

\clearpage

\section{Coherent control mechanism}
\noindent
As established in Figs.~2 and 3 in the main text, the effective field acting on the Er spin depends not only on the Er-Ti distance, but also on the direction of the Er-Ti bond. 
This indicates that the spin interaction should be described by an anisotropic exchange tensor whose value depends on both the relative position of the two atoms and their local orbital configurations. 
The static Hamiltonian for spin-spin interaction is:
\begin{equation}
    \mathcal{H}_{\rm int}
    =
    \boldsymbol{S}_{\rm Ti}
    \mathcal{J}
    \boldsymbol{S}_{\rm Er},
    \label{eq:H_static_exchange}
\end{equation}
where
\begin{equation}
    \mathcal{J}
    =
    \mathcal{J}
    \left(
    \boldsymbol{r},
    \boldsymbol{\eta}_{\rm Ti},
    \boldsymbol{\eta}_{\rm Er}
    \right).
    \label{eq:J_static_def}
\end{equation}
Here $\boldsymbol{r}$ is the vector connecting the Ti and Er atoms, while $\boldsymbol{\eta}_{\rm Ti}$ and $\boldsymbol{\eta}_{\rm Er}$ characterize the orientations of the Ti and Er ground-state orbitals with respect to the lattice.

Under RF excitation, the microscopic configuration of the Er-Ti pair is perturbed. We describe this perturbation as a time-dependent infinitesimal trajectory $X_{\rm RF}(t)$ in the configuration space $\boldsymbol{q} = \left( \boldsymbol{r}, \boldsymbol{\eta}_{\rm Ti}, \boldsymbol{\eta}_{\rm Er} \right)$.
Because $\mathcal{J}$ is a rank-2 tensor acting in spin space, an RF-induced rotation of the local spin/orbital frames changes not only its components but also the spin basis in which those components are defined. 
Its modulation is therefore described by a covariant, or Lie-derivative, change rather than by an ordinary directional derivative alone.
The complete first-order variation is given by the Lie derivative of $\mathcal{J}$ along the RF-induced trajectory,
\begin{equation}
    \delta \mathcal{J}(t)
    =
    \mathcal{L}_{X_{\rm RF}(t)}\mathcal{J}.
    \label{eq:Lie_derivative_start}
\end{equation}
Consequently, the most general RF driving Hamiltonian to first order in the RF amplitude is
\begin{equation}
    \mathcal{H}_{\rm d}(t)
    =
    \boldsymbol{S}_{\rm Ti}
    \left[
    \mathcal{L}_{X_{\rm RF}(t)}\mathcal{J}
    \right]
    \boldsymbol{S}_{\rm Er}.
    \label{eq:H_drive_Lie}
\end{equation}
The RF perturbation can change the exchange tensor in two physically distinct ways. 
First, it can modify the tensor components by changing the bond geometry or the local
orbital configuration. Second, it can rotate the spin frames in which the tensor indices
are defined, creating off-diagonal terms in $\mathcal{J}(\boldsymbol{q}(t))$. The Lie derivative provides a compact way to keep track of both effects:
\begin{equation}
    \mathcal{L}_{X_{\rm RF}(t)}\mathcal{J}
    =
    \underbrace{X_{\rm RF}(t)\cdot\nabla\mathcal{J}}_
    {\text{component change from geometry/orbital modulation}}
    +
    \underbrace{
    A_{\rm Ti}^{\mathsf T}(t)\mathcal{J}
    +
    \mathcal{J} A_{\rm Er}(t)}
    _{\text{rotation of the Ti and Er spin indices}} .
    \label{eq:Lie_decomposition}
\end{equation}
Here \(A_{\rm Ti}(t)\) and \(A_{\rm Er}(t)\) are the corresponding
infinitesimal generators that rotate the Ti and Er spin frames. 
The first term is
therefore the ordinary directional derivative of \(\mathcal{J}\) along the RF trajectory,
whereas the second term accounts for the fact that \(\mathcal{J}\) is a rank-two tensor
whose two spin indices transform with the local Ti and Er frames.
\subsection{Modulation of diagonal components in $\mathcal{J}$}
\noindent
The first term in Eq.~\eqref{eq:Lie_decomposition} can be written as
\begin{equation}
    X_{\rm RF}(t)\cdot\nabla\mathcal{J}
    =
    \delta\boldsymbol{r}(t)\cdot\nabla_{\boldsymbol r}\mathcal{J}
    +
    \delta\boldsymbol{\eta}_{\rm Ti}(t)\cdot
    \nabla_{\boldsymbol{\eta}_{\rm Ti}}\mathcal{J}
    +
    \delta\boldsymbol{\eta}_{\rm Er}(t)\cdot
    \nabla_{\boldsymbol{\eta}_{\rm Er}}\mathcal{J} .
    \label{eq:ordinary_derivative_terms}
\end{equation}
In Eq.~\eqref{eq:ordinary_derivative_terms}, the first term describes a piezoelectric modulation of the Er-Ti bond, whereas the second terms describe direct RF-induced changes of the local orbital configurations which are determined by the coupling between the microwave and the atomic orbitals~\cite{reina2025efficient,maeda2005microwave}.
The piezoelectric contribution is in general a weak effect~\cite{phark2023electric,reina2025efficient}, and therefore we neglect this term. 
In addition, within a minimal description, the direct orbital-deformation terms mainly modulate the diagonal values of $\mathcal{J}$ ($\mathcal{J}_{xx}$, $\mathcal{J}_{yy}$ and $\mathcal{J}_{zz}$), and therefore they do not mix spin indexes.

\begin{figure}[htbp]
    \centering
    \includegraphics[width = 1\textwidth]{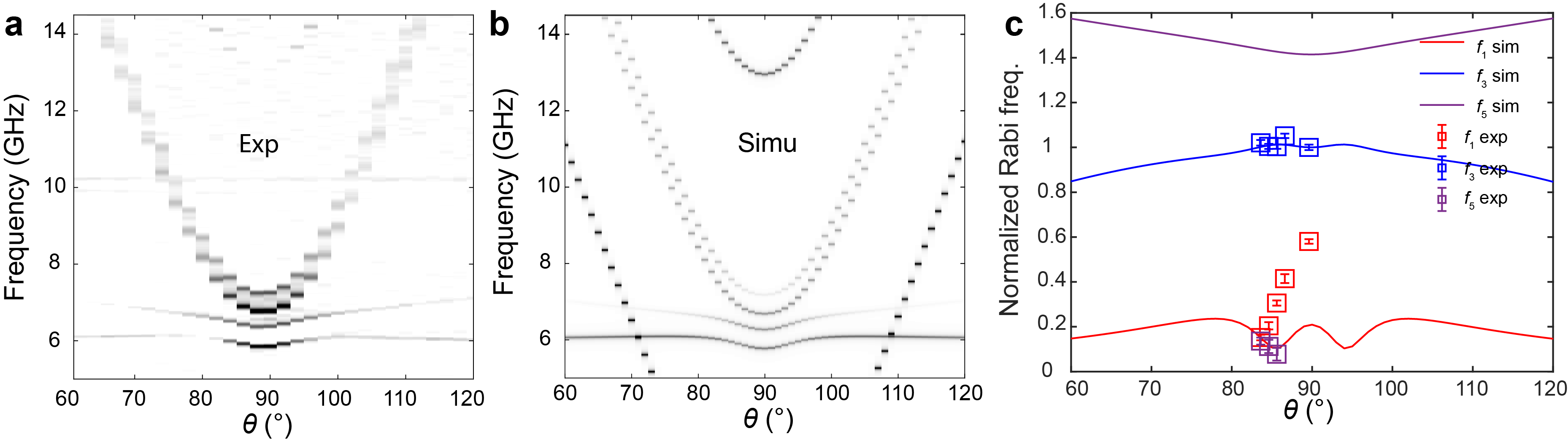}
    \caption{\textbf{Modulation of the $\boldsymbol{S}_{\rm Ti}^x\delta\mathcal{J}_{xx}\boldsymbol{S}_{\rm Er}^x$ term in the (3,0) pair.} \textbf{a,} Experimental ESR map as a function of the polar angle $\theta$ ($\phi = 52\degree$, $V_{\rm DC} = 100$ mV, $I_{\rm DC} = 20$ pA, $V_{\rm RF} = 20$ mV). \textbf{b,} Simulated ESR map as a function of the polar angle $\theta$ ($\phi = 52\degree$). \textbf{c,} The normalized Rabi frequencies as a function of polar angle (normalized by the Rabi frequencies of $f_3$ at $\phi = 52\degree$ and $\theta = 90\degree$). The solid lines are simulations while the colored squares are the experimental results. All experimental data in (\textbf{c}) were acquired with $V_{\rm DC} = 100$ mV, $I_{\rm DC} = 20$ pA, $V_{\rm RF} = 100$ mV, $\boldsymbol{B}_{\rm ext} = 0.25$ T and $\phi = 52\degree$.}
    \label{FigS_Modulating_Jxx}
\end{figure}

\begin{figure}[htbp]
    \centering
    \includegraphics[width = 0.8\textwidth]{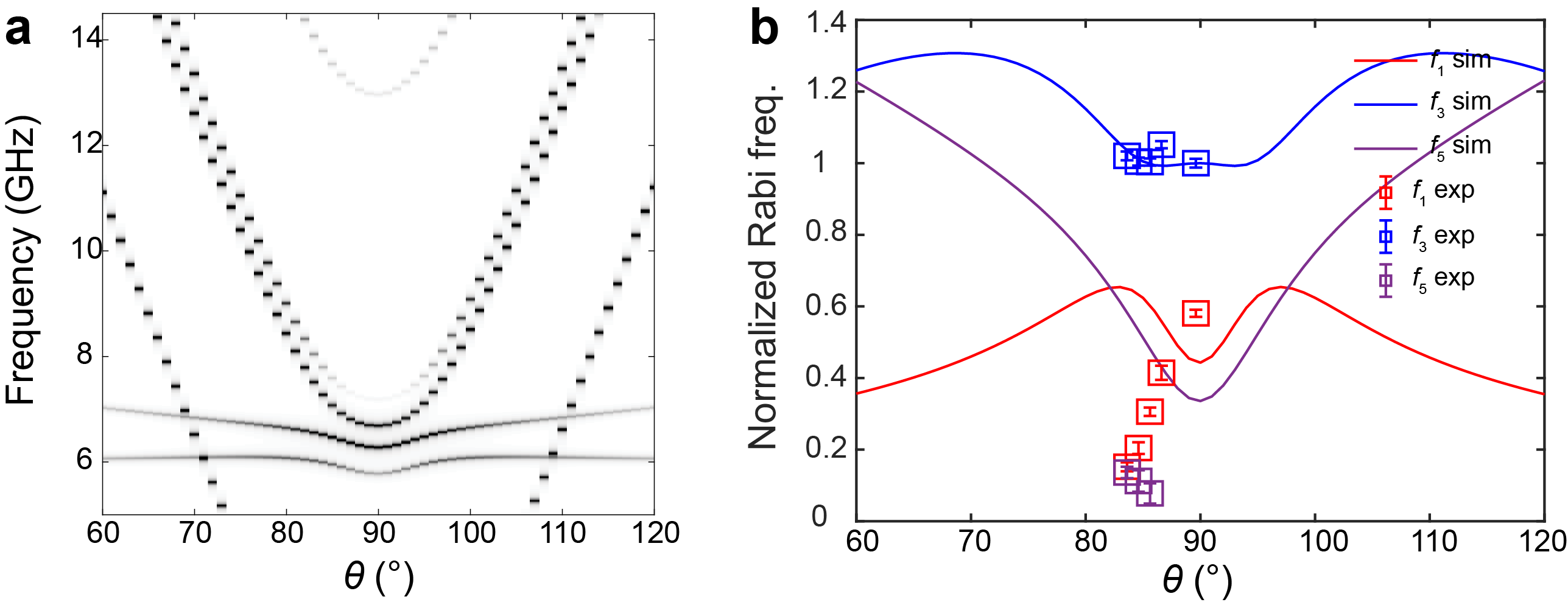}
    \caption{\textbf{Modulation of the $\boldsymbol{S}_{\rm Ti}^y\delta\mathcal{J}_{yy}\boldsymbol{S}_{\rm Er}^y$ term in the (3,0) pair.} \textbf{a,}  Simulated ESR map as a function of the polar angle $\theta$ ($\phi = 52\degree$). \textbf{b,} The normalized Rabi frequencies as a function of polar angle (normalized by the Rabi frequencies of $f_3$ at $\phi = 52\degree$ and $\theta = 90\degree$). The solid lines are simulation while the colored squares are the experimental results. All experimental data were acquired with $V_{\rm DC} = 100$ mV, $I_{\rm DC} = 20$ pA, $V_{\rm RF} = 100$ mV, $\boldsymbol{B}_{\rm ext} = 0.25$ T and $\phi = 52\degree$.}
    \label{FigS_Modulating_Jyy}
\end{figure}

\begin{figure}[htbp]
    \centering
    \includegraphics[width = 0.8\textwidth]{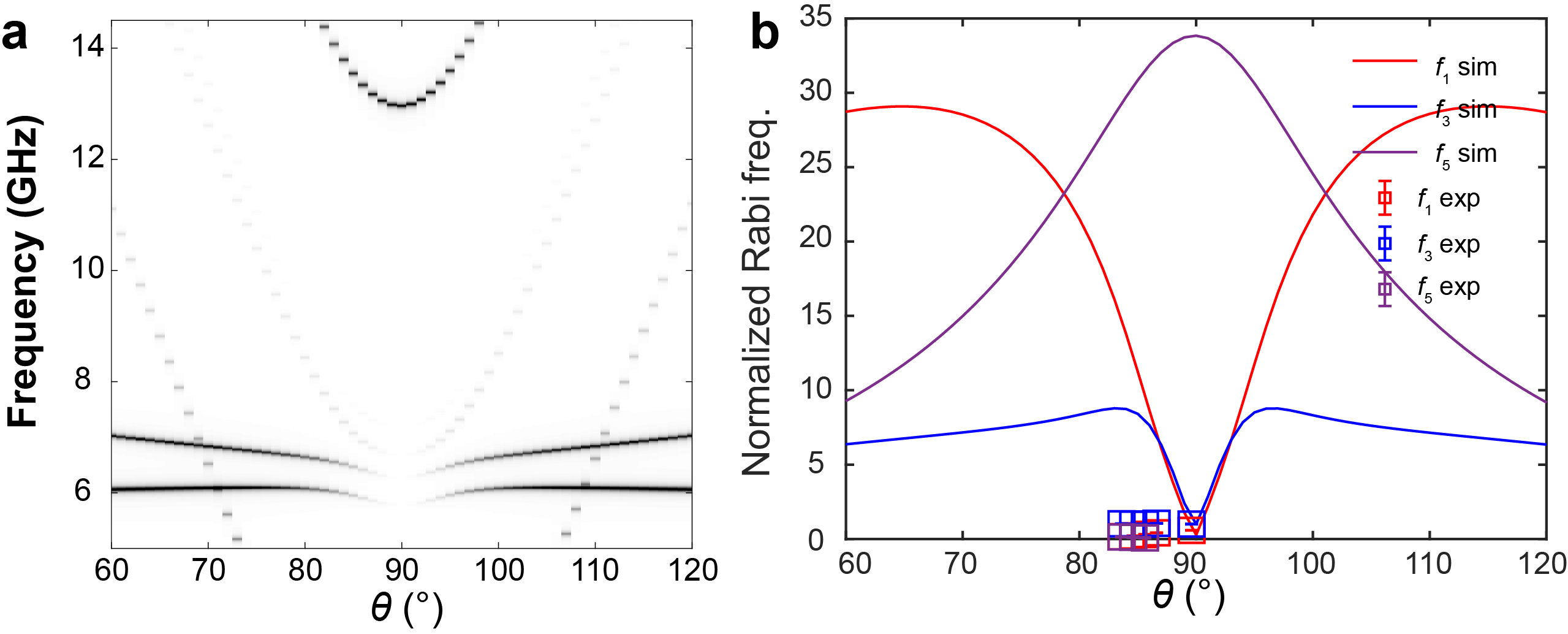}
    \caption{\textbf{Modulation of the $\boldsymbol{S}_{\rm Ti}^z\delta\mathcal{J}_{zz}\boldsymbol{S}_{\rm Er}^z$ term in the (3,0) pair.} \textbf{a,}  Simulated ESR map as a function of the polar angle $\theta$ ($\phi = 52\degree$). \textbf{b,} The normalized Rabi frequencies as a function of polar angle (normalized by the Rabi frequencies of $f_3$ at $\phi = 52\degree$ and $\theta = 90\degree$). The solid lines are simulation while the colored squares are the experimental results. All experimental data were acquired with $V_{\rm DC} = 100$ mV, $I_{\rm DC} = 20$ pA, $V_{\rm RF} = 100$ mV, $\boldsymbol{B}_{\rm ext} = 0.25$ T and $\phi = 52\degree$.}
    \label{FigS_Modulating_Jzz}
\end{figure}

To investigate the driving term involving the spin-index-preserving terms: i.e. $\boldsymbol{S}_{\rm Ti}^x\delta\mathcal{J}_{xx}\boldsymbol{S}_{\rm Er}^x$, $\boldsymbol{S}_{\rm Ti}^y\delta\mathcal{J}_{yy}\boldsymbol{S}_{\rm Er}^y$ and $\boldsymbol{S}_{\rm Ti}^z\delta\mathcal{J}_{zz}\boldsymbol{S}_{\rm Er}^z$, we calculate the ESR spectrum within linear response theory by treating the RF excitation as a weak time-dependent perturbation: $\mathcal{H}_{\rm drive} =\boldsymbol{S}_{\rm Ti}^x\delta\mathcal{J}_{xx}\cos{(\omega_{\rm RF} t)}\boldsymbol{S}_{\rm Er}^x$, $\mathcal{H}_{\rm drive} =\boldsymbol{S}_{\rm Ti}^y\delta\mathcal{J}_{yy}\cos{(\omega_{\rm RF} t)}\boldsymbol{S}_{\rm Er}^y$ and $\mathcal{H}_{\rm drive} =\boldsymbol{S}_{\rm Ti}^z\delta\mathcal{J}_{zz}\cos{(\omega_{\rm RF} t)}\boldsymbol{S}_{\rm Er}^z$
added to the static spin Hamiltonian $\mathcal{H}$. 
After diagonalizing $\mathcal{H}$, the absorption intensity at frequency $\omega_{\rm RF}$ is obtained by summing over all allowed transitions between eigenstates \(\ket{n}\) and \(\ket{m}\),
\begin{equation}
    I(\omega) \propto
    \sum_{n,m}
    \left(p_n-p_m\right)
    \left|\bra{m}{\mathcal{H}_{\rm drive}}\ket{n}\right|^2
    L\!\left[\omega-\frac{E_m-E_n}{\hbar}\right],
\end{equation}
where \(p_n \propto \exp(-E_n/k_{\rm B}T)\) is the Boltzmann population and \(L(\omega)\) is a Lorentzian lineshape accounting for the finite linewidth, which we set to be 10 MHz. 
This approach directly connects the calculated ESR peak positions to the eigenenergy differences of $\mathcal{H}$, while the peak intensities are determined by the transition matrix elements of the RF driving operator and the population difference between two levels. 
Here we neglect the tip polarization for the readout process, which does not qualitatively change the results.
To compare with the experimental ESR map and Rabi frequencies as a function of polar angle $\theta$ (with $\phi$ fixed at 52$\degree$, Fig.~\ref{FigS_Modulating_Jxx}\textbf{a}), we simulate ESR map and normalized Rabi frequencies (normalized by $f_3$ at $\theta = 90\degree$ ) by modulating the  $\boldsymbol{S}_{\rm Ti}^x\delta\mathcal{J}_{xx}\boldsymbol{S}_{\rm Er}^x$, $\boldsymbol{S}_{\rm Ti}^y\delta\mathcal{J}_{yy}\boldsymbol{S}_{\rm Er}^y$ and $\boldsymbol{S}_{\rm Ti}^z\delta\mathcal{J}_{zz}\boldsymbol{S}_{\rm Er}^z$ terms.
The ESR map and the normalized Rabi frequencies are presented in Figs.~\ref{FigS_Modulating_Jxx}\textbf{b-c}, \ref{FigS_Modulating_Jyy}\textbf{a-b}  and \ref{FigS_Modulating_Jzz}\textbf{a-b}.
The simulation clearly fails to reproduce the key experimental observations:
\begin{itemize}
\item For the case of modulating $\boldsymbol{S}_{\rm Ti}^x\delta\mathcal{J}_{xx}\boldsymbol{S}_{\rm Er}^x$ (as in Fig.~\ref{FigS_Modulating_Jxx}), the ESR intensity as well as the normalized Rabi frequencies are too strong compared with experimental results. In addition, the ESR intensity of $f_3$ shows almost no change when $\theta$ approaches $90\degree$, while the experimental results show a clear decrease of ESR intensity when the field is aligned in the in-plane direction.
\item For the case of modulating $\boldsymbol{S}_{\rm Ti}^y\delta\mathcal{J}_{yy}\boldsymbol{S}_{\rm Er}^y$ (as in Fig.~\ref{FigS_Modulating_Jyy}), the ESR intensity of $f_4$ decreases when field goes in-plane, and the $f_3$ ESR peak shows no clear decrease when field goes in plane. More importantly, the normalized Rabi frequencies for DQT are twice as strong as in the experimental results.
\item For the case of modulating $\boldsymbol{S}_{\rm Ti}^z\delta\mathcal{J}_{zz}\boldsymbol{S}_{\rm Er}^z$ (as in Fig.~\ref{FigS_Modulating_Jzz}), the ESR peak intensities of $f_1$ to $f_5$ show all distinctly different trends compared with experimental results in Fig.~\ref{FigS_Modulating_Jxx}\textbf{a} and the normalized Rabi frequencies cannot capture the experimental results.
\end{itemize}

\subsection{Modulating the off-diagonal term through rotational modulation}
\noindent
Since all the terms in Eq.~\eqref{eq:ordinary_derivative_terms} fail to account for the experimental observation, we therefore focus on the second part of Eq.~\eqref{eq:Lie_decomposition}, namely the spin-index rotation of the anisotropic exchange tensor: $\mathcal{H}_{\rm rot}= A_{\rm Ti}^{\mathsf T}(t)\mathcal{J}+ \mathcal{J} A_{\rm Er}(t)$, which describes the local basis change when traveling along the trajectory and creates off-diagonal components in $\mathcal{J}(\boldsymbol{q}(t))$.
Note that for a pure translational trajectory in the space of $\boldsymbol{q}$, $\mathcal{H}_{\rm rot}$ is zero, since it does not change the local spin basis of Er and Ti.
We therefore consider the rotational trajectory. For a finite rotation, this tensorial part of the Lie derivative can be represented as
\begin{equation}
    \mathcal{J}
    \rightarrow
    \mathcal{J}_{\rm rot}(t)
    =
    R_{\rm Ti}^{\mathsf T}(t)
    \mathcal{J}
    R_{\rm Er}(t),
    \label{eq:J_rot_general}
\end{equation}
where $R_{\rm Ti}(t)$ and $R_{\rm Er}(t)$ are $SO(3)$ rotation matrices acting on the Ti and Er spin indices. The corresponding RF-induced modulation of the exchange tensor is therefore
\begin{equation}
    \delta\mathcal{J}_{\rm rot}(t)
    =
    R_{\rm Ti}^{\mathsf T}(t)
    \mathcal{J}
    R_{\rm Er}(t)
    -
    \mathcal{J}.
    \label{eq:deltaJ_rot_general}
\end{equation}
The associated driving Hamiltonian is based on the rotational modulation of $\mathcal{J}$:
\begin{equation}
    \mathcal{H}_{\rm rot}(t)
    =
    \boldsymbol{S}_{\rm Ti}
    \delta\mathcal{J}_{\rm rot}(t)
    \boldsymbol{S}_{\rm Er}.
    \label{eq:H_rot_general}
\end{equation}

\begin{figure}[htbp]
    \centering
    \includegraphics[width = 0.65\textwidth]{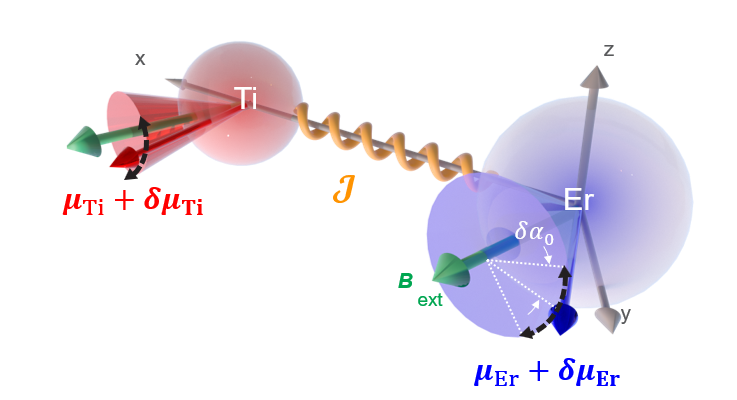}
    \caption{\textbf{Wobbling the magnetic moment around the external magnetic field.} The rotational modulation of $\mathcal{J}$ can be viewed in the rotating frame as the wobbling of magnetic moment around the magnetic field $\boldsymbol{B}_{\rm ext}$. The change of Ti magnetic moment $\delta\boldsymbol{\mu}_{\rm Ti}$ will exert an oscillating field on the Er spin which drives the Er spin transitions. And similarly, the wobbling of the Er spin exerts an oscillating field on the Ti spin.}
    \label{FigS_wobble_3D}
\end{figure}

To minimize the energetic cost of the RF excitation, we consider the rotational mode of the local anisotropy frame that leaves the dominant Zeeman energy unchanged. Physically, the RF electric field modulates the orbital structure that defines the principal axes of the $\mathfrak{g}$-tensor and the exchange tensor $\mathcal{J}^{\rm exc}$. Through spin-orbit coupling, this modulation is equivalently described as an apparent wobbling of the magnetic moments around the external field $\boldsymbol{B}_{\rm ext}$ (Fig.~\ref{FigS_wobble_3D}). 
Since this wobbling changes only the azimuthal direction of $\boldsymbol{\mu}_{i}$ while preserving its projection onto $\boldsymbol{B}_{\rm ext}$, the Zeeman energy is unchanged to leading order.
For this energetically favorable mode, we first write Eq~\eqref{eq:H_rot_general} in the form of magnetic moment ($\boldsymbol{\mu}_{\rm Er(Ti)}=\mu_{\rm B}\mathfrak{g}_{\rm Er(Ti)}\cdot\boldsymbol{S}$):
\begin{equation}
    \label{eq:H_rot_moment}
    \mathcal{H}_{\rm rot}=\boldsymbol{\mu}_{\rm Ti}\mathcal{J}^{\rm M}\boldsymbol{\mu}_{\rm Er}
\end{equation}
with $\mathcal{J}^{\rm M}=\mathfrak{g}^{-1}_{\rm Ti}\mathcal{J}\mathfrak{g}^{-1}_{\rm Er}/\mu_{\rm B}^2$. In this framework, we write:
\begin{equation}
    R_{\rm Ti}(t)
    =
    R_{\rm Er}(t)
    =
    R_{B}\!\left[\delta\alpha(t)\right],
    \label{eq:common_rotation}
\end{equation}
where $R_{B}$ is a rotation around $\boldsymbol{B}_{\rm ext}$, and
\begin{equation}
   \delta\alpha(t)
    =
    \delta\alpha_0\cos(\omega_{\rm RF}t)
    \label{eq:dphi_harmonic}
\end{equation}
is the RF-induced rotational modulation. Equation~\eqref{eq:deltaJ_rot_general} then becomes
\begin{equation}
    \delta\mathcal{J}^{\rm M}_{\rm rot}(t)
    =
    R_{B}^{\mathsf T}\!\left[\delta\alpha(t)\right]
    \mathcal{J}^{\rm M}
    R_{B}\!\left[\delta\alpha(t)\right]
    -
    \mathcal{J}^{\rm M}.
    \label{eq:deltaJ_RB}
\end{equation}
For small $\delta\alpha$, the rotation matrix can be expanded as
\begin{equation}
    R_{B}\!\left[\delta\alpha(t)\right]
    =
    I  +
    \delta\alpha(t)G_{B}+ \mathcal{O}\!\left[\delta\alpha(t)^2\right],
    \label{eq:RB_expand}
\end{equation}
where $G_{B}$ is the generator of rotations around $\hat{\boldsymbol{n}}_{B}$. Since $G_{B}$ is antisymmetric,
\begin{equation}
    G_{B}^{\mathsf T}
    =
    -G_{B}.
    \label{eq:LB_antisymmetric}
\end{equation}
Substituting Eq.~\eqref{eq:RB_expand} into Eq.~\eqref{eq:deltaJ_RB}, we obtain
\begin{align}
    \delta\mathcal{J}^{\rm M}_{\rm rot}(t)
    &=
    \left[
    I-\delta\alpha(t)G_{B}
    \right]
    \mathcal{J}^{\rm M}
    \left[
    I+\delta\alpha(t)G_{B}
    \right]
    -
    \mathcal{J}^{\rm M}
    +
    \mathcal{O}\!\left[\delta\alpha(t)^2\right]
    \nonumber \\
    &=\delta\alpha(t)\left(\mathcal{J}^{\rm M}G_{B} -  G_{B}\mathcal{J}^{\rm M} \right)
    +
\mathcal{O}\!\left[\delta\alpha(t)^2\right].
    \label{eq:deltaJ_commutator_derivation}
\end{align}
Thus, to leading order,
\begin{equation}
    \delta\mathcal{J}^{\rm M}_{\rm rot}(t)
    =
    \delta\alpha(t)
    \left[
    \mathcal{J}^{\rm M},G_{B}
    \right],
    \label{eq:deltaJ_commutator}
\end{equation}
The rotational driving Hamiltonian is therefore:
\begin{equation}
    \mathcal{H}_{\rm rot}(t)
    = \mathcal{H}_{\rm 1}(t)=
    \delta\alpha(t)
    \boldsymbol{\mu}_{\rm Ti}
    \left[
   \mathcal{J}^{\rm M},G_{B}
    \right]
    \boldsymbol{\mu}_{\rm Er}.
    \label{eq:H_rot_final}
\end{equation}
Equation~\eqref{eq:H_rot_final} shows that the RF drive is controlled by the commutator between the anisotropic exchange tensor and the generator of rotations around the external magnetic field.
Note Equation~\eqref{eq:H_rot_final} is a generalized form of Eq.~2 in the main text that accounts for the $\mathfrak{g}$-tensor anisotropy of both spins. 
This term vanishes for an isotropic exchange interaction, $\mathcal{J}^{\rm M}=J I$, because in that case
\begin{equation}
    \left[
    \mathcal{J}^{\rm M},G_B
    \right]
    =
    0.
\end{equation}
Therefore, the anisotropy of $\mathcal{J}$ is essential for this driving mechanism.

The same result can be understood in the co-moving anisotropy frame. 
In this frame, the orbital landscape is static, while the magnetic moments appear to counter-wobble around $\boldsymbol{B}_{\rm ext}$ (as shown in Fig.~\ref{FigS_wobble_3D}):
\begin{equation}
    \mathcal{H}_{J}(t)
    =
    \left[
    R_B\!\left(\delta\alpha(t)\right)
    \boldsymbol{\mu}_{\rm Ti}
    \right]^{\mathsf T}
    \mathcal{J}^{\rm M}
    \left[
    R_B\!\left(\delta\alpha(t)\right)
    \boldsymbol{\mu}_{\rm Er}
    \right].
    \label{eq:wobbling_frame}
\end{equation}
Equivalently, in spin-coordinate space this corresponds to the $\mathfrak{g}$-stretched rotations $W_i(t)=\mathfrak{g}_{i}^{-1}R_B(t)\mathfrak{g}_{i}$. Thus the ``spin wobbling'' picture is not a literal direct rotation of the spin by the microwave field, but an equivalent description of the RF-induced modulation of the orbital anisotropy frame.

\begin{figure}[htbp]
    \centering
    \includegraphics[width = 1\textwidth]{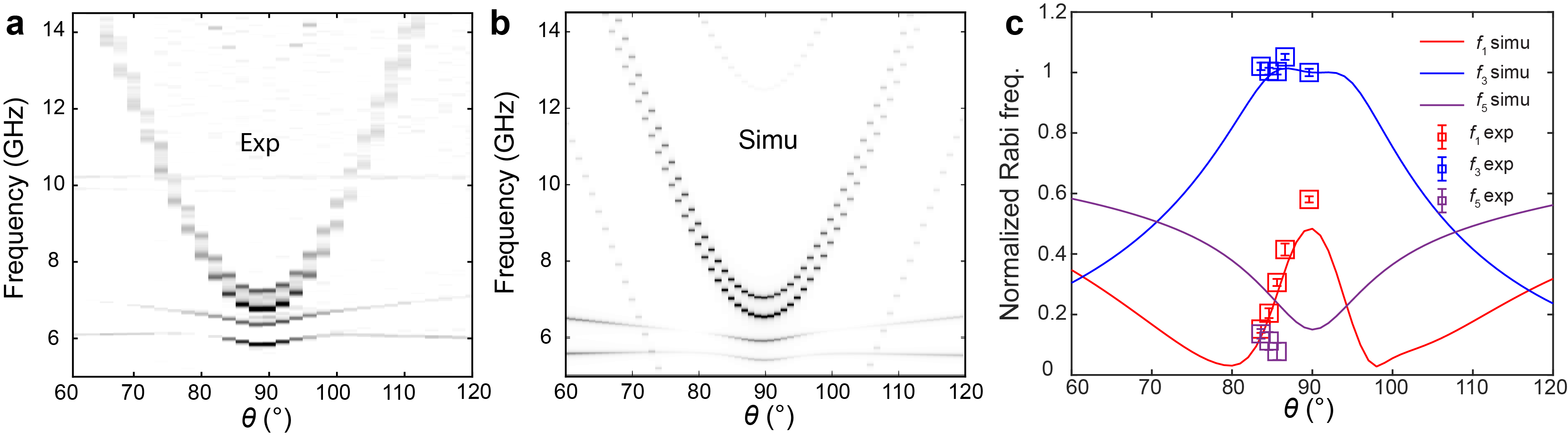}
    \caption{\textbf{Simulation based on rotational modulation for the (3, 0) pair.} \textbf{a,} Experimental ESR map as a function of the polar angle $\theta$ ($\boldsymbol{B}_{\rm ext} = 0.25$ T, $\phi = 52\degree$, $V_{\rm DC} = 100$ mV, $I_{\rm DC} = 20$ pA, $V_{\rm RF} = 20$ mV). \textbf{b,} Simulated ESR map as a function of the polar angle $\theta$ ($\phi = 52\degree$) assuming a wobbling angle of $\delta\alpha_0 = 0.1\pi$. \textbf{c,} The normalized Rabi frequencies as a function of the polar angle (normalized by the Rabi frequencies of $f_3$ at $\phi = 52\degree$ and $\theta = 90\degree$). The solid lines are simulations while the colored squares are the experimental results. All experimental data in (\textbf{c}) were acquired with $V_{\rm DC} = 100$ mV, $I_{\rm DC} = 20$ pA, $V_{\rm RF} = 100$ mV, $\boldsymbol{B}_{\rm ext} = 0.25$ T, $\phi = 52\degree$.}
    \label{FigS_wobble}
\end{figure}

\begin{figure}[htbp]
    \centering
    \includegraphics[width = 1\textwidth]{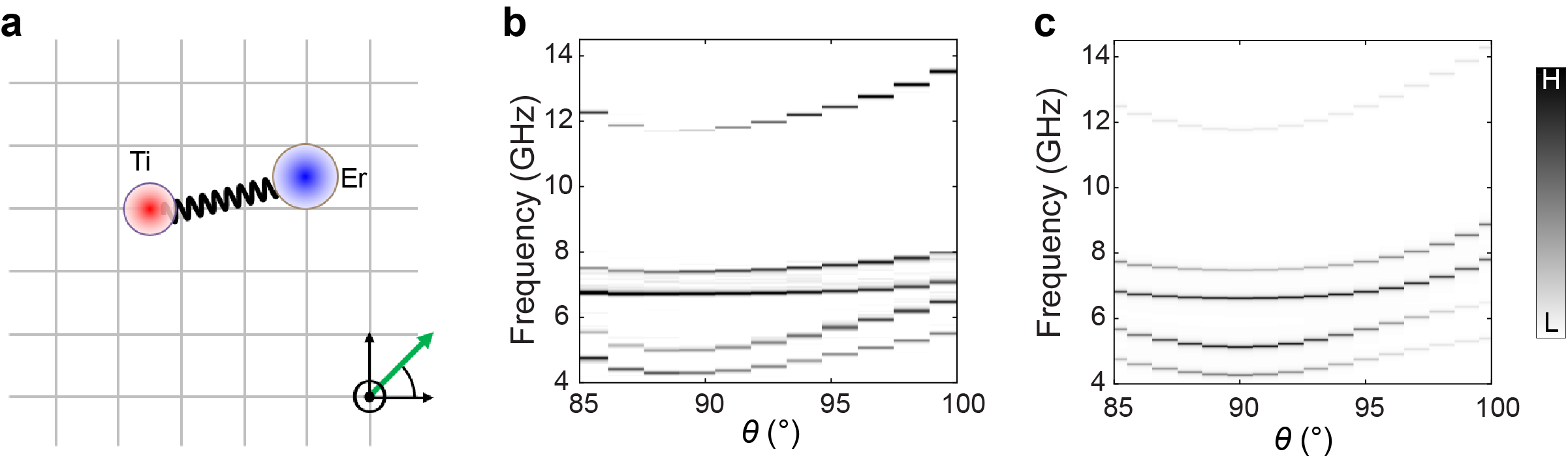}
    \caption{\textbf{Simulation based on rotational modulation for the (2.5, $-$0.5) pair.}
\textbf{a,} Schematic of the (2.5, $-$0.5) Er-Ti spin pair.
\textbf{b,} Experimental ESR map as a function of the polar angle $\theta$ for the (2.5, $-$0.5) pair with $\phi$ fixed at $52\degree$ ($V_{\rm DC} = 100$ mV, $I_{\rm DC} = 20$ pA, $V_{\rm RF} = 20$ mV).
\textbf{c,} Simulated ESR map as a function of the polar angle $\theta$ for the (2.5, $-$0.5) pair ($\phi = 52\degree$) assuming a wobbling angle of $\delta\alpha_0 = 0.1\pi$. All experimental data were acquired under the conditions of $\boldsymbol{B}_{\rm ext} = 0.25$ T, $\phi = 52\degree$.}
    \label{FigS_wobble2}
\end{figure}

We simulate the ESR map using Eq.~\eqref{eq:H_rot_final} and compare it with experimental results.
In Figs.~\ref{FigS_wobble} and ~\ref{FigS_wobble2}, we present the experimental and simulated ESR maps as a function of the polar angle $\theta$, with $\phi$ fixed at $52\degree$, for the (3,0) and (2.5, $-$0.5) Er-Ti spin pairs. 
In both cases, the experimental ESR maps exhibit clear and systematic trends as the polar angle is varied:
\begin{itemize}
\item For both the (3,0) and (2.5, $-$0.5) pairs, the DQT intensity diminishes as the polar angle approaches the in-plane direction ($90\degree$).
\item For the (3,0) pair, the ESR transition intensities from $f_1$ to $f_4$ increase as the magnetic field approaches the in-plane orientation.
\end{itemize}
The simulated ESR maps based on our model of rotational modulation of the spin-spin coupling tensor reproduce these trends well, as shown in Fig.~\ref{FigS_wobble}\textbf{b} and Fig.~\ref{FigS_wobble2}\textbf{c}.
Furthermore, for the (3,0) pair, the model also captures the angular dependence of the Rabi frequencies, as presented in Fig.~\ref{FigS_wobble}\textbf{c} where we show the Rabi frequencies, normalized to the $f_3$ Rabi frequency at $\theta = 90\degree$, as a function of the polar angle $\theta$ with $\phi$ fixed at $52\degree$.


\clearpage

\subsection{Contribution of dipolar interaction}
\begin{figure}[htbp]
    \centering
    \includegraphics[width = 1\textwidth]{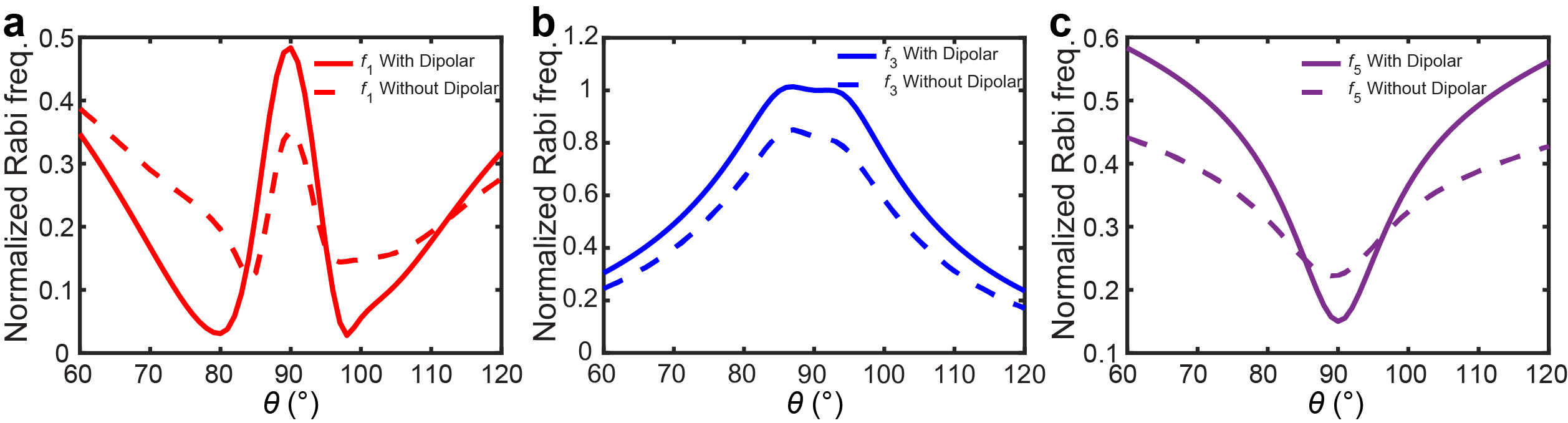}
    \caption{\textbf{Contribution of the dipolar interaction to the Rabi frequency.}
\textbf{a-c,} Normalized Rabi frequencies calculated with (solid lines) and without (dashed lines) the dipolar interaction as a function of the polar angle $\theta$ for $f_1$ (\textbf{a}), $f_3$ (\textbf{b}), and $f_5$ (\textbf{c}). All simulations were performed under an external magnetic field of $\boldsymbol{B}_{\rm ext} = 0.25$ T with $\phi = 52\degree$ and a wobbling angle of $\delta\alpha_0 = 0.1\pi$.}
    \label{FigS_Dip}
\end{figure}

\noindent
In this subsection, we evaluate the contribution of the dipolar interaction to the Rabi frequencies of $f_1$, $f_3$, and $f_5$ (normalized to the $f_3$ Rabi frequency at $\theta = 90\degree$, with $\phi$ fixed at $52\degree$). As shown in Fig.~\ref{FigS_Dip}, the exclusion of the dipolar interaction does not qualitatively modify the angular dependence of the Rabi frequencies as a function of the polar angle $\theta$.

\clearpage

\subsection{Special cases of coherent control mechanism: $\theta = 90\degree$, $\phi = 0\degree$ or $\phi = 90\degree$}
\begin{figure}[htbp]
    \centering
    \includegraphics[width = 0.8\textwidth]{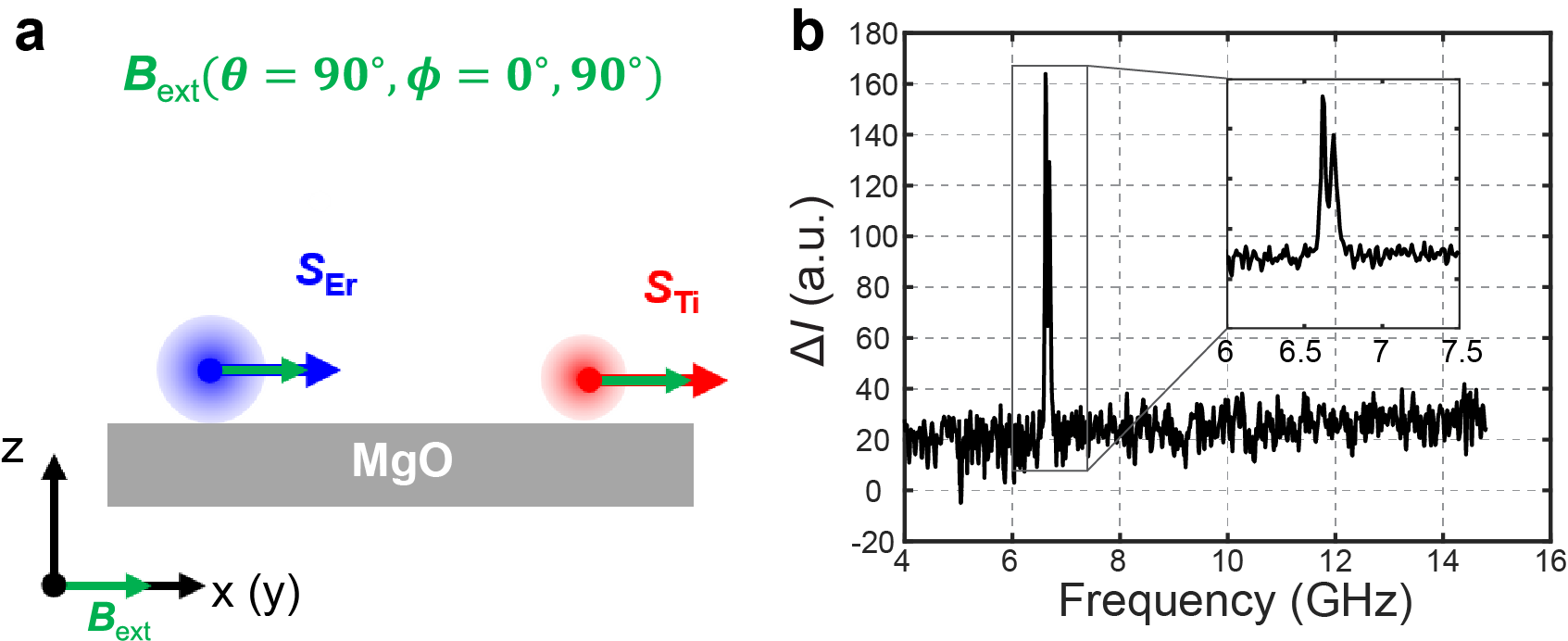}
    \caption{\textbf{Special cases of the coherent control mechanism with $\boldsymbol{B}_{\rm ext}$ aligned along the $\textbf{\textit{x}}$ or $\textbf{\textit{y}}$-axis.}
\textbf{a,} Schematic illustration. When the magnetic field is aligned along an in-plane axis, both spins align with the field and therefore with each other.
\textbf{b,} ESR spectrum measured at $\phi = 0\degree$ and $\theta = 90\degree$ ($\boldsymbol{B}_{\rm ext} = 0.3$ T, $V_{\rm DC} = 50$ mV, $I_{\rm DC} = 20$ pA, $V_{\rm RF} = 20$ mV). Only Ti transitions are observed.}
    \label{FigS_Sce1}
\end{figure}

\noindent
As shown in Fig.~\ref{FigS_Sce1}\textbf{a}, when the magnetic field is aligned along the $x$- or $y$-axis, both the Er and Ti spins are aligned with the field. In this configuration, rotational modulation around the magnetic-field direction does not modify the spin orientations and therefore does not produce any driving.
Under these conditions, only the Ti transitions are detected as shown in Fig.~\ref{FigS_Sce1}\textbf{b}.

\subsection{Special cases of coherent control mechanism: $\theta = 90\degree$, $0\degree<\phi<90\degree$ }
\begin{figure}[htbp]
    \centering
    \includegraphics[width = 0.4\textwidth]{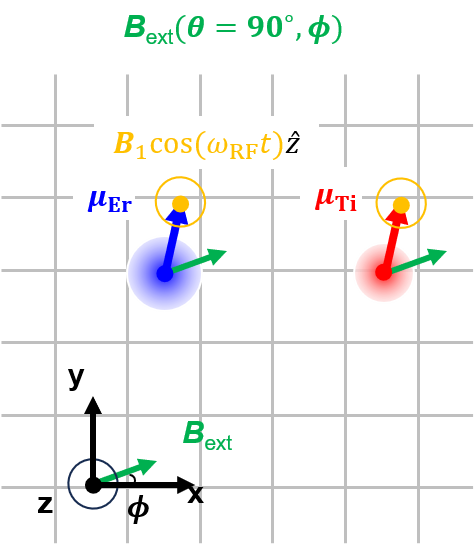}
    \caption{\textbf{Special scenario of coherent control mechanism with $\boldsymbol{B}_{\rm ext}$ aligned in-plane $\theta$ = 90$\degree$ with $0\degree<\phi<90\degree$.}}
    \label{FigS_Sce2}
\end{figure}

\noindent
When the magnetic field is aligned in-plane ($\theta = 90\degree$) but with a finite azimuthal angle $\phi$, the Er and Ti spins preferentially align closer to the $y$ axis due to the in-plane $\mathfrak{g}$-tensor anisotropies.
In this configuration, the wobbling of each magnetic moment around the magnetic-field direction induces a spin-vector variation with an out-of-plane component, as indicated by the yellow circle in Fig.~\ref{FigS_Sce2}. 
This generates an out-of-plane driving field $\boldsymbol{B}_1$.
The resulting $\boldsymbol{B}_1$ couples to the large out-of-plane $\mathfrak{g}$-tensor component $\mathfrak{g}^{\rm Er}_{zz}$ of the Er spin ($\mathfrak{g}^{\rm Er}_{zz} = 9.59$), which is substantially larger than the in-plane components. 
Consequently, this mechanism leads to the fast Rabi frequencies observed experimentally.

\clearpage
\section{Readout process in Er-Ti pair}
\noindent
In this section, we discuss the readout process for the ESR measurement in Er(B)-Ti pair.
In the case of Er(O)-Ti pair, where Ti and Er(O) spin have a lifetime imbalance~\cite{reale2024electrically} so that the spin pumping effect is taken into account for the read-out process.
However, in the Er(B)-Ti pair, both spins have a comparable lifetime, on the order of 50 ns, yet one can still read out the Er spin states, even when Er and Ti have a substantial Zeeman energy difference (see Fig.~5 of the main text).
Here we show that because of this rotational modulation of spin-spin interaction, even when two spins are in the Zeeman product state, one can still read out the Er-spin population change from Ti.
This is because rotational modulation of spin interaction as shown by the driving Hamiltonian $\mathcal{H}_1$ (Eq.~\eqref{eq:H_rot_final}) involves driving both spins simultaneously.

To simulate the ESR readout from the Ti spin in the coupled Er-Ti spin pair, we solve a Markovian master equation in Lindblad form for the driven open quantum system~\cite{gorini1976completely,lindblad1976generators}. 
The static Hamiltonian includes the anisotropic Zeeman terms of the two spins in the external magnetic field together with their mutual interaction, using the Hamiltonian defined throughout this work (Eq.~1 of the main text). 
The numerical simulations are implemented using the Quantum Toolbox in Python (QuTiP)~\cite{johansson2012qutip}.

To incorporate relaxation and thermalization, we construct global thermal collapse operators using the strategy from previous studies on similar platforms~\cite{le2026overcoming,broekhoven2024protocol}
with the relaxation rates given by the $T_1$ of the Er and Ti spins measured from experiments (Section VII). 
In the absence of the RF drive, these dissipative terms relax the system toward the undriven thermal steady state at the chosen temperature $T$ = 1.0 K. 
This undriven steady state is used as the initial density matrix for the continuous-wave ESR simulations.
The driving Hamiltonian is given by Eq.\eqref{eq:H_rot_final}.
For each frequency $f_{\rm RF}$, we solve the Lindblad equation in the time domain, discard the initial transient evolution, and then average the expectation value of the readout operator over the subsequent periodic steady-state regime.
The readout operator is the tip polarization $\hat{\boldsymbol{P}}_{\rm tip}$, which is assumed to align with the magnetic field in accordance with the assumption that the tip field lies parallel or antiparallel to the external magnetic field.
Repeating this procedure over the full frequency sweep yields the simulated ESR spectrum. 
To highlight the resonant contribution, we subtract an off-resonant background estimated from the spectral edges and normalize the resulting signal. 
The ESR peak positions therefore correspond to drive frequencies that resonantly enhance the change in the sensor readout, while the peak amplitudes reflect the strength of the driven response under the combined action of coherent driving and thermal dissipation.

\begin{figure}[htbp]
    \centering
    \includegraphics[width = 0.6\textwidth]{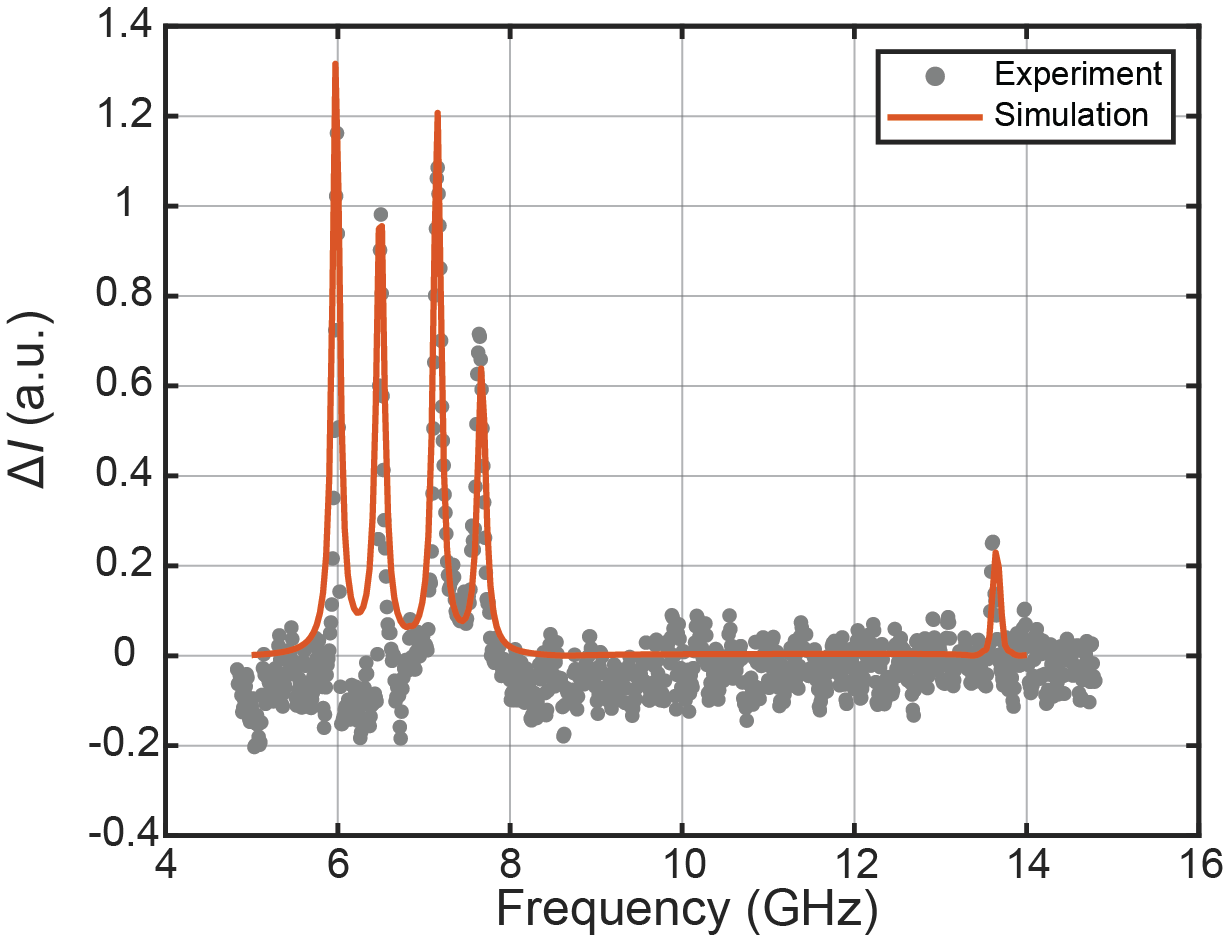}
    \caption{\textbf{Simulated vs. experimental ESR spectra.} The gray circles represent the experimental data, while the red solid line represents the simulated ESR spectrum obtained using QuTiP. The readout operator is chosen as the tip polarization operator $\hat{\boldsymbol{P}}_{\rm tip}$, and the tip driving term is taken as $0.17\,\boldsymbol{S}_{{\rm Ti},x}$, which is chosen to reproduce the peak heights of the Ti transitions ($f_1$ and $f_2$). $\boldsymbol{B}_{\rm ext}=0.25~\mathrm{T}$, $\theta=84.5\degree$, and $\phi=52\degree$.}
    \label{Readout_ExpVsSim}
\end{figure}

In Fig.~\ref{Readout_ExpVsSim}, we present the experimental and simulated ESR spectra using the model described above. 
The tip field is obtained from the fitting procedure described in Section III and the tip driving is modeled by $0.17\,\boldsymbol{S}_{{\rm Ti},x}$ to fit the experimental Ti transition peak heights ($f_1$ and $f_2$ in Fig.~\ref{Readout_ExpVsSim}).

\clearpage
\section{Simulation of Er(O)-Ti pair}
\begin{figure}[htbp]
    \centering
    \includegraphics[width = 1\textwidth]{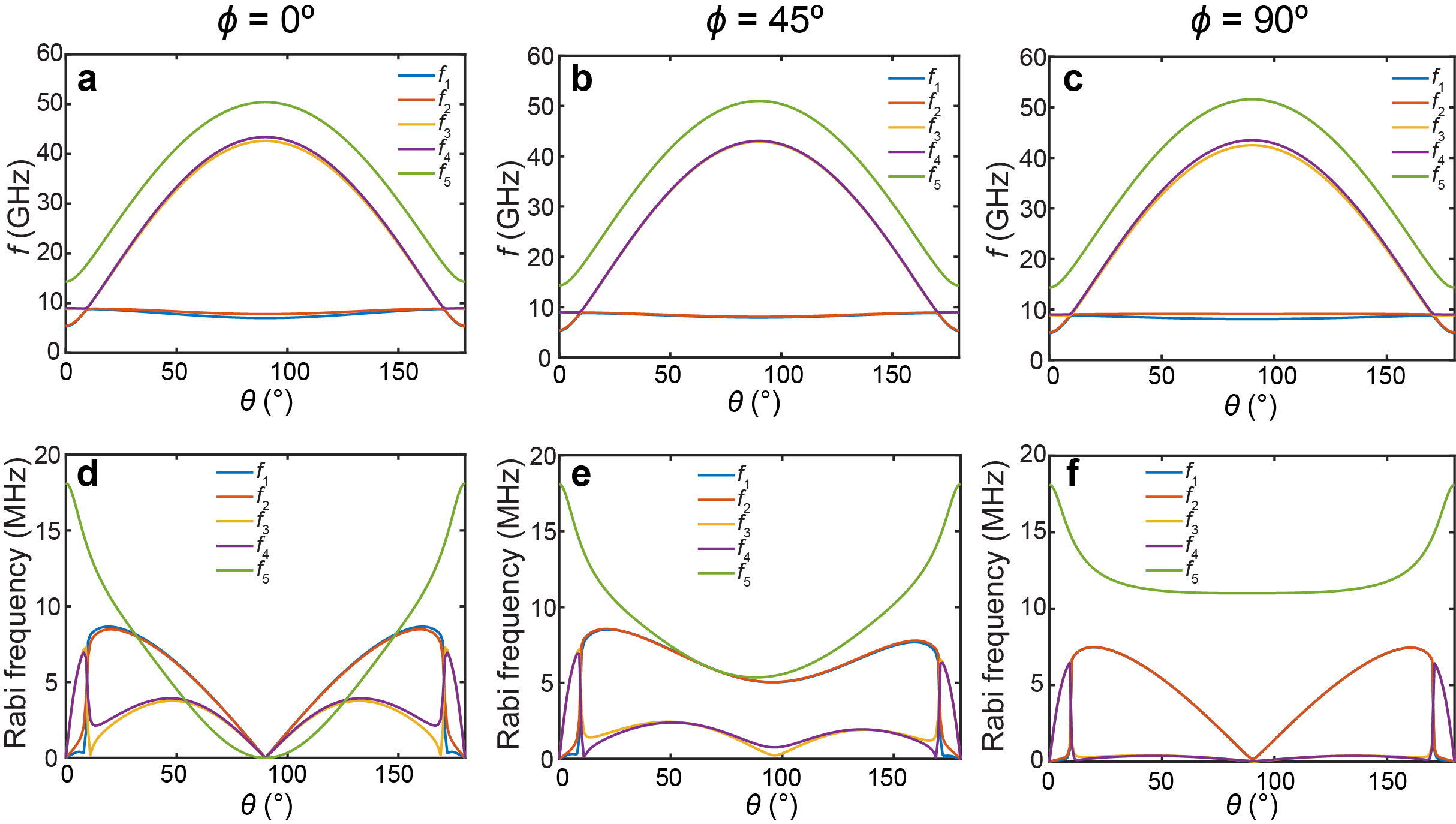}
    \caption{\textbf{Simulation of Rabi frequencies and ESR transitions for the Er(O)-Ti pair in previous work~\cite{reale2024electrically}.} \textbf{a-c,} ESR transition for Er(O)-Ti pair as a function of polar angle $\theta$ for azimuthal angle of $\phi $ = 0$\degree$ (\textbf{a}),  45$\degree$ (\textbf{b}), 90$\degree$ (\textbf{c}). \textbf{d-f,} Rabi frequency for Er(O)-Ti pair as a function of polar angle $\theta$ for azimuthal angle of $\phi $ = 0$\degree$ (\textbf{d}),  45$\degree$ (\textbf{e}), 90$\degree$ (\textbf{f}). The simulation is done under the condition of $\boldsymbol{B}_{\rm ext}$ = 0.32 T as in the literature and we assume a wobbling angle of $\delta\alpha_0$ = 0.1$\pi$. }
    \label{FigS_ErO}
\end{figure}

\noindent
In this section, we apply our proposed mechanism to the Er(O)-Ti pair, where the Er atom is adsorbed on an oxygen binding site. 
As reported in previous work~\cite{reale2024electrically}, the spin-spin interaction is $J_0 = 48$ MHz.
In the effective spin-$1/2$ description obtained by projecting onto the ground-state doublet, the principal values of the $\mathfrak{g}$-tensor are $\mathfrak{g}_{ xx}^{\rm Er(O)} = 9.6$, $\mathfrak{g}_{yy}^{\rm Er(O)} = 9.6$, and $\mathfrak{g}_{zz}^{\rm Er(O)} = 1.2$~\cite{reale2024electrically}.
The simulated ESR transitions and Rabi frequencies as a function of the polar angle $\theta$ for different values of $\phi$ are presented in Fig.~\ref{FigS_ErO}, assuming the same wobbling angle of $\delta\alpha_0 = 0.1\pi$.
Comparing the Rabi frequencies of the Er transitions ($f_3$ and $f_4$ in Fig.~\ref{FigS_ErO}) with those of the Ti transitions ($f_1$ and $f_2$ in Fig.~\ref{FigS_ErO}), we observe that the Ti transitions exhibit higher Rabi frequencies.

\begin{figure}[htbp]
    \centering
    \includegraphics[width = 0.5\textwidth]{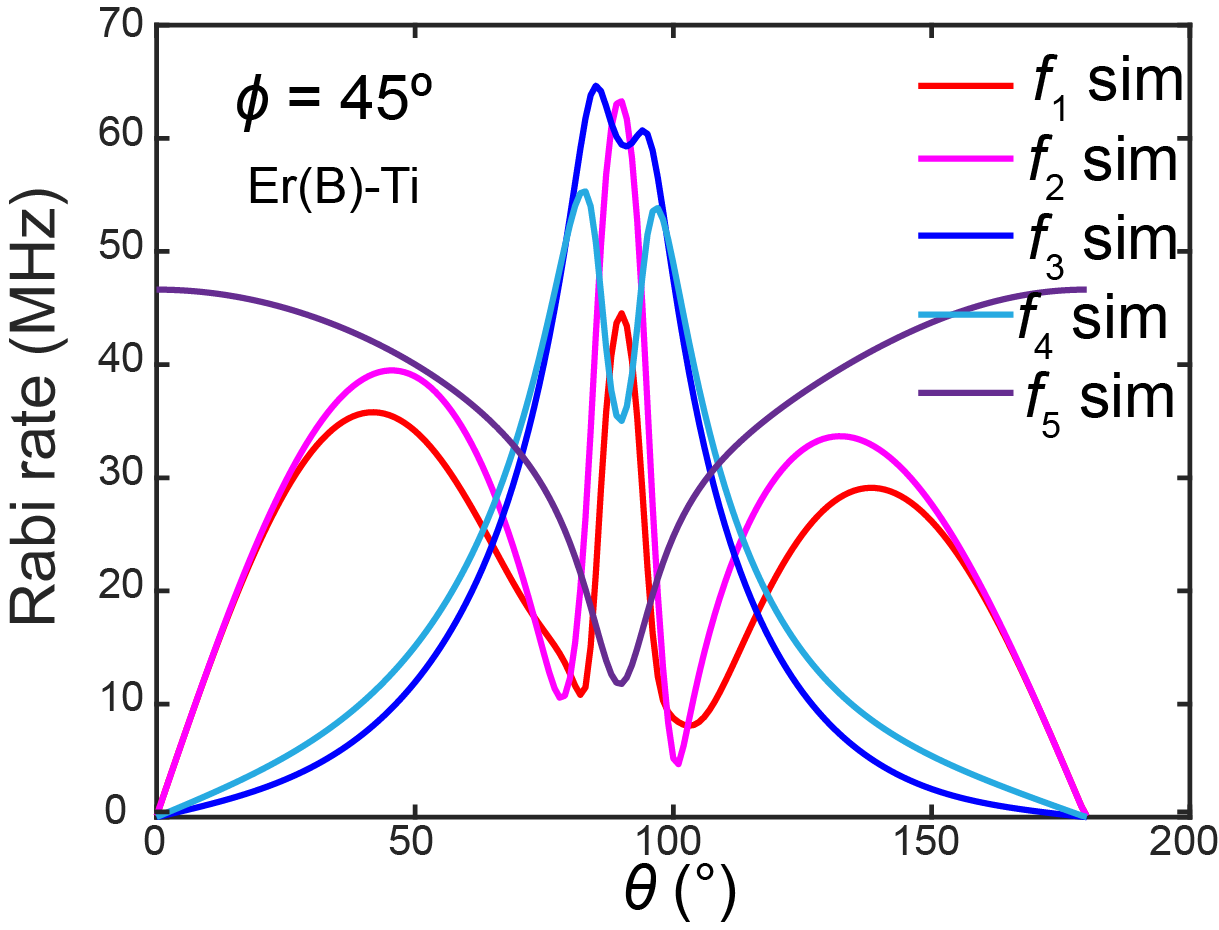}
    \caption{\textbf{Simulation of Rabi frequencies for the (3, 0) Er(B)-Ti pair.}  Rabi frequency for (3, 0) Er(B)-Ti pair as a function of polar angle $\theta$. The simulation is done under the condition of $\boldsymbol{B}_{\rm ext}$ = 0.32 T and $\phi $ = 45$\degree$ with $\delta\alpha_0$ = 0.1$\pi$. }
    \label{FigS_ErB}
\end{figure}

Using the same rotational modulation model with the same wobbling angle $\delta\alpha_0$ = 0.1$\pi$, we compare the simulated Rabi frequencies between the Er(B)-Ti (3, 0) pair (Fig.~\ref{FigS_ErB}) and the Er(O)-Ti pair (Fig.~\ref{FigS_ErO}(\textbf{e})) with the azimuthal angle $\phi$ = 45$\degree$ and $\boldsymbol{B}_{\rm ext}$ = 0.32 T.
The Rabi frequencies for the Er(O)-Ti pair under the same condition are around 2 MHz as shown in Fig.~\ref{FigS_ErO}(\textbf{e}), which are $\sim$30 times smaller than the case of the Er(B)-Ti (3, 0) under the same condition.

\end{document}